\documentclass[aps,pra,twocolumn]{revtex4-1}
\usepackage[T1]{fontenc}
\usepackage[latin1]{inputenc}
\usepackage{lmodern}
\expandafter\let\csname equation*\endcsname\relax
\expandafter\let\csname endequation*\endcsname\relax
\usepackage{amsmath}
\usepackage{amssymb}
\usepackage{graphicx}
\usepackage{xcolor}
\usepackage{natbib}
\usepackage{bm}
\usepackage{bbold}
\usepackage[mathscr]{euscript}
\usepackage[cal=boondoxo]{mathalfa}
\usepackage{comment}
\renewcommand\footnotemark{}

\def\ii{{\rm i}}  
\def\GG{{\bf G}}
\def\ggm{{\bf g}}  
 
\def\db{{\bf d}}  
\def\Eb{{\bf E}}  
\def\rb{{\bf r}}  
\def\fb{{\bf f}}  
 
\def\Omegab{{\bf \Omega}}  
\def\pb{{\bf p}}

\def\ge{\sigma_{ge}}  
\def\eg{\sigma_{eg}}  
\def\hge{\hat{\sigma}_{ge}}  
\def\heg{\hat{\sigma}_{eg}}  
\def\db{\textbf{d}}

\def\ga{\Gamma_{\rm 1D}}

\def\oa{\omega_{\rm A}}

\def\bra#1{\mathinner{\langle{#1}|}}
\def\ket#1{\mathinner{|{#1}\rangle}}
\def\braket#1{\mathinner{\langle{#1}\rangle}}

\begin{document}

\title{Atom-light interactions in quasi-1D nanostructures: a Green's function perspective}
\author{A. Asenjo-Garcia$^{1,2,\dagger}$}
\email[E-mail: ]{ana.asenjo@caltech.edu}
\author{ J. D. Hood$^{1,2}$}
\thanks{These authors contributed equally to this research}
\author{D. E. Chang$^3$}
\author{H. J. Kimble$^{1,2}$}
\affiliation{$^1$ Norman Bridge Laboratory of Physics MC12-33, California Institute of Technology, Pasadena, CA 91125, USA}
\affiliation{$^2$ Institute for Quantum Information and Matter, California Institute of Technology, Pasadena, CA 91125, USA}
\affiliation{$^3$ ICFO-Institut de Ciencies Fotoniques, The Barcelona Institute of Science and Technology, 08860 Castelldefels (Barcelona), Spain}

\date{\today}
\begin{abstract}
Based on a formalism that describes atom-light interactions in terms of the classical electromagnetic Green's function, we study the optical response of atoms and other quantum emitters coupled to one-dimensional photonic structures, such as cavities, waveguides, and photonic crystals. We demonstrate a clear mapping between the transmission spectra and the local Green's function, identifying signatures of dispersive and dissipative interactions between atoms. We also demonstrate the applicability of our analysis to problems involving three-level atoms, such as electromagnetically induced transparency. Finally we examine recent experiments, and anticipate future observations of atom-atom interactions in photonic bandgaps.
\end{abstract}
\pacs{42.50.Ct, 42.50.Nn}
\maketitle

\section{Introduction}
As already noticed by Purcell in the first half of the past century, the decay rate of an atom can be either diminished or enhanced by tailoring its dielectric environment \cite{P1946,K1981,HK1989}. Likewise, by placing more than one atom in the vicinity of photonic nanostructures, one can curtail or accelerate their collective decay. In addition to modifying the radiative decay, nanophotonic structures can be employed to spatially and spectrally engineer atom-light interactions, thus obtaining fundamentally different atom dynamics to those observed in free-space \cite{LMS15}. 

In the past decade, atoms and other quantum emitters have been interfaced with the electromagnetic fields of a plethora of quasi-1D nanostructured reservoirs, ranging from high-quality optical  \cite{VY03,YSH04, ADW06, HBW07, HSC10,TTL13} and microwave \cite{PSB11,RGP12} cavities to dielectric \cite{BHB09,BHK04,LVB09,VRS10,GCA12,VSJ14,polzik,CGC16}, metallic \cite{CSD07,AMY07,HKS11,GMM11}, and superconducting \cite{LFL13,DS13} waveguides. Photonic crystal waveguides, periodic dielectric structures that display a bandgap where light propagation is forbidden \cite{JMW95,Y1987}, have been proposed as promising candidates to study long- and tunable-range coherent interactions between quantum emitters \cite{J1987,JW1990, K1990,DHH15}. Due to the different character of the guided modes at various frequencies within the band structure of the photonic crystal, the interaction of the quantum emitters with the nanostructure can be remarkably distinct depending on the emitter resonance frequency. Far away from the bandgap, where light propagates, the guided modes resemble those of a conventional waveguide. Close to the bandgap, but still in the propagating region, the fields are similar to those of a quasi-1D cavity, whereas inside the bandgap the fields become evanescent, decaying exponentially. 

All these regimes have been recently explored in the lab, where atoms \cite{GHH15,GHY14,YHM14} and quantum dots \cite{LSJ08,YTB15,LMS15} have been interfaced with photonic crystal waveguides. Most of these experiments have been performed in conditions where the resonance frequency of the emitter lies outside the bandgap. However, very recently, the first experiments of atoms \cite{HGA16} and superconducting qubits \cite{LH16} interacting with evanescent modes in the bandgap of photonic crystal waveguides have been reported.

Within this context, it has become a necessity to understand the rich spectral signatures of atom-like emitters interacting through the guided modes of quasi one-dimensional nanophotonic structures within a unified framework that extends beyond those of cavity \cite{MS1990} or waveguide QED \cite{SF05}. In this work, we employ a formalism based on the classical electromagnetic Green's function \cite{GW96,DKW02,BW07,DSF10,BI12} to characterize the response of atoms that interact by emitting and absorbing photons through the guided mode of the nanostructure. Since the fields in the vicinity of the structure might have complex spatial and polarization patterns, the full Green's function is only known analytically for a handful of systems (such as planar multilayer stacks \cite{T95}, infinite nanofibers \cite{KD04,KGB05,QBJ16}, and a few more \cite{BW07}) and beyond that one has to resort to numerical solvers of Maxwell's equations. However, in quasi-1D nanostructures, one can isolate the most relevant guided mode and build a simple prescription for the 1D Green's function that accounts for the behavior of this mode, greatly simplifying the problem.

In the first part of the article, we summarize the procedure to obtain an effective atom-atom Hamiltonian, in which the guided-mode fields are effectively eliminated and the atom interactions are written in terms of Green's functions \cite{GW96,DKW02,BW07}. We then apply this formalism to a collection of atoms in different quasi one-dimensional dielectric environments, and analyze the atomic transmission and reflection spectra in terms of the eigenvalues of the matrix consisting of the Green's functions between every pair of atoms. We show that, in the linear (low-saturation) regime, asymmetry in the transmission spectra and frequency shifts are signatures of coherent atom-light interactions, whereas symmetric lineshapes reveal dissipation. We also tackle the problem of three-level atoms coupled to quasi-1D nanostructures under electromagnetically induced transparency (EIT) conditions, and derive analytical expressions for the polaritonic band structure of such systems in terms of the eigenvalues of the Green's function matrix. Finally, based on the rapid technical advances in fabrication of both optical and microwave structures, we project observable signatures that can be made in the next generation of experiments of atoms and superconducting qubits interacting in the bandgap of photonic crystal waveguides. 

\section{Atom-light interactions in terms of Green's functions}
Much effort has gone into developing a quantum formalism to describe atoms coupled to radiation. A conventional technique is to express the field in terms of a set of eigenmodes of the system, with corresponding creation and annihilation operators $a^{\dagger}$ and $a$ \cite{SF05}.  This canonical quantization technique is well suited for approximately closed systems such as high-Q cavities and homogeneous structures such as waveguides, both of which have simple eigenmode decompositions.  However, the application of this quantization scheme to more involved nanostructures is not straight forward.  Further, the formalism is not suited for dispersive and absorbing media as the commutation relations for the field operators are not conserved \cite{BS06}. 

Instead, here we describe atom-light interactions using a quantization scheme based on the classical electromagnetic Green's function, valid for any medium characterized by a linear and isotropic dielectric function $\epsilon(\rb,\omega)$, closely following the work of Welsch and colleagues \cite{GW96,DKW02,BW07,BI12}.  In the following, we employ this formalism to derive an atom-atom Hamiltonian in which the field is effectively eliminated, yielding an expression that only depends on atomic operators. Moreover, once the dynamics of the atoms is solved, the electric field at every point along the quasi one-dimensional structure can be recovered through an expression that relates the field to the atomic operators.

Classically, the field $\Eb(\rb,\omega)$ at a point $\rb$ due to a source current $\textbf{j}(\rb',\omega)$ at $\rb'$ is obtained by means of the propagator of the electromagnetic field, the dyadic Green's function (or Green's tensor), as $\Eb(\rb,\omega)=\ii\mu_0\omega\int d\rb' \,\GG(\rb,\rb',\omega)\cdot \textbf{j}(\rb',\omega)$. In particular, for a dipole source $\pb$ located at $\rb_0$, the current is $\textbf{j}(\rb,\omega)=-\ii\omega \pb\,\delta(\rb-\rb_0)$, and the field reads $\Eb(\rb,\omega)=\mu_0\omega^2\,\GG(\rb,\rb_0,\omega)\cdot\pb$. The tensorial structure of the Green's function accounts for the vectorial nature of the electromagnetic field, as a dipole directed along the $\hat{x}$-direction can create a field polarized not only along $\hat{x}$, but also along $\hat{y}$ and $\hat{z}$. Throughout this manuscript, the Green's tensor will be also denoted as Green's function.

The Green's function $\GG(\rb,\rb',\omega)$ is the fundamental solution of the electromagnetic wave equation, and obeys \cite{NH06}:
\begin{align}\label{gf}
\bm{\nabla}\times\bm{\nabla}\times \GG(\rb,\rb',\omega)-\frac{\omega^2}{c^2}\epsilon(\rb,\omega)\, \GG(\rb,\rb',\omega)=\delta(\rb-\rb')\mathbb{1},
\end{align}
where $\epsilon(\rb,\omega)$ is the medium relative permittivity. For a scalar permittivity, Lorentz reciprocity holds and, then, $\GG^{\rm T}(\rb,\rb',\omega)=\GG(\rb',\rb,\omega)$, where T stands for transpose (and operates on the polarization indexes). This formalism ignores the possibility of systems made non-reciprocal by the material response \cite{BBS12} (e.g., non-symmetric permittivity tensors), although it does in principle cover the interesting case of chiral atom-light interactions \cite{SMH15,LMS16}, where the time-reversal symmetry breaking is due to the atomic states.

In analogy to its classical counterpart, the electric field operator at frequency $\omega$ can be written in terms of bosonic annihilation (creation) operators $\hat{\textbf{f}}$ ($\hat{\textbf{f}}^{\dagger}$) as \cite{GW96}
\begin{widetext}
\begin{align}\label{field}
\hat{\Eb}(\rb,\omega)&=\ii\mu_0\,\omega^2 \sqrt{\frac{\hbar\epsilon_0}{\pi}}\int d\rb'\,\sqrt{\text{Im}\{\epsilon(\rb',\omega)\}}\, \text{\GG}(\rb,\rb'\omega)\cdot\hat{\fb} (\rb',\omega)+ \text{h.c.}=\hat{\Eb}^+(\rb,\omega)+\hat{\Eb}^-(\rb,\omega),
\end{align}
\end{widetext}
where $\hat{\Eb}^{+(-)}(\rb,\omega)$ is the positive (negative) frequency component of the field operator, h.c. stands for Hermitian conjugate, and the total field operator reads $\hat{\Eb}(\rb)=\int d\omega \,\hat{\Eb}(\rb,\omega)$. Within this quantization framework, $\hat{\textbf{f}}(\rb,\omega)$ is associated with the degrees of freedom of local material polarization noise, which accompanies the material dissipation $\text{Im}\{\epsilon(\rb,\omega)\}$ as required by the fluctuation-dissipation theorem \cite{BI12}. This expression guarantees the fulfillment of the canonical field commutation relations, even in the presence of material loss. The appearance of the Green's function reveals that the \textit{quantumness} of the system is encoded in either the correlations of the noise operators $\hat{\textbf{f}}$ or in any other quantum sources (such as atoms), but the field propagation obeys the wave equation and as such the spatial profile of the photons is determined by the classical propagator. 

We  now want to investigate the evolution of $N$ identical two-level atoms of resonance frequency $\oa$ that interact through a guided mode probe field of frequency $\omega_{\rm p}$. Within the Born-Markov approximation, we trace out the photonic degrees of freedom, obtaining an effective atom-atom Hamiltonian \cite{DKW02,CJG12,CST14}. This approximation is valid when the atomic correlations decay much slower than the photon bath correlations, or, in other words, when the Green's function is characterized by a broad spectrum, which can be considered to be flat over the atomic linewidth. Then, the atomic density matrix $\hat{\rho}_{\rm A}$ evolves according to $\dot{\hat{\rho}}_{\rm A}=-(\ii/\hbar)\,[\mathcal{H},\hat{\rho}_{\rm A}]+\mathcal{L}[\hat{\rho}_{\rm A}]$ \cite{MS1990}. Within the rotating wave approximation, and in the frame rotating with the probe field frequency, the Hamiltonian and Lindblad operators read 
\begin{subequations}
\begin{align}\label{ham}
\mathcal{H}=&-\hbar\Delta_{\rm A}\sum_{i=1}^N\hat{\sigma}_{ee}^i-\hbar\sum_{i,j=1}^N J^{ij}\hat{\sigma}_{eg}^i\hat{\sigma}_{ge}^j\\\nonumber
&-\sum_{i=1}^N\left(\db\cdot\hat{\Eb}_{\rm p}^- (\rb_i)\,\hat{\sigma}_{ge}^i+\db^*\cdot\hat{\Eb}_{\rm p}^+(\rb_i)\,\hat{\sigma}_{eg}^i\right),
\end{align}
\begin{equation}\label{lind}
\mathcal{L}[\hat{\rho}_{\rm A}]=\sum_{i,j=1}^N\frac{\Gamma^{ij}}{2}\,\left(2\hat{\sigma}_{ge}^i\hat{\rho}_{\rm A}\hat{\sigma}_{eg}^j-\hat{\sigma}_{eg}^i\hat{\sigma}_{ge}^j\hat{\rho}_{\rm A}-\hat{\rho}_{\rm A}\hat{\sigma}_{eg}^i\hat{\sigma}_{ge}^j\right),
\end{equation}
\end{subequations}
where $\hat{\Eb}_{\rm p}$ is the guided mode probe field, and $\Delta_{\rm A}=\omega_{\rm p}-\omega_{\rm A}$ is the detuning between the guided mode probe field and the atom. The dipole moment operator is expressed in terms of the dipole matrix elements as  $\hat{\pb}_j = \db^* \, \hat{\sigma}^j_{eg} + \db \, \hat{\sigma}^j_{ge} $, where $\heg^j=\ket{e}\bra{g}$ is the atomic coherence operator between the ground and excited states of atom $j$, and  $\db=\braket{g|\hat{\pb}_j |e}$ is the dipole matrix element associated with that transition. The spin-exchange and decay rates are 
\begin{subequations}
\begin{equation}
J^{ij}=(\mu_0\omega_{\rm p}^2/\hbar)\,\db^*\cdot\text{Re}\,\mathbf{G}(\rb_i,\rb_j,\omega_{\rm p})\cdot\db,
\end{equation}
\begin{equation}
\Gamma^{ij} =(2\mu_0\,\omega_{\rm p}^2/\hbar)\,\db^*\cdot\text{Im}\,\mathbf{G}(\rb_i,\rb_j,\omega_{\rm p})\cdot\db.
\end{equation}
\end{subequations}

Note that the dispersive and dissipative atom-atom couplings are given in terms of the total Green's function of the medium. For a given dielectric geometry, $\mathbf{G}(\rb_i,\rb_j,\omega_{\rm p})$ can be calculated either numerically or analytically to obtain quantitative predictions for the spin exchange and decay matrix elements. On the other hand, given some basic assumptions about the regime of interest, one can construct a simple effective model for $\mathbf{G}(\rb_i,\rb_j,\omega_{\rm p})$. This enables one to broadly capture a number of physical systems and gain general insight, and we take this approach here.

In particular, we assume that there is a single 1D guided band to which the atoms predominantly couple, and explicitly separate its contribution from the Green's function, $\GG(\rb_i,\rb_j,\omega_{\rm p})=\GG_{\rm 1D}(\rb_i,\rb_j,\omega_{\rm p})+\GG'(\rb_i,\rb_j,\omega_{\rm p})$. The second term in the sum contains the atom-atom interactions mediated by all other free space and guided modes. We further assume that only collective interactions via the explicitly separated 1D channel are important, while $\GG'(\rb_i,\rb_j,\omega_{\rm p})$ only provides independent single-atom decay and energy shifts. Then, we can write $J^{ij}=J^{ij}_{\rm 1D}+J'\delta_{ij}$ and $\Gamma^{ij}=\Gamma^{ij}_{\rm 1D}+\Gamma'\delta_{ij}$, where $\delta_{ij}$ is the Kronecker delta. In particular, in free-space, $\Gamma'$ is simply $\Gamma_0=(2\mu_0\,\omega_{\rm p}^2/\hbar)\,\db^*\cdot\text{Im}\,\mathbf{G}_0(\rb_i,\rb_i,\omega_{\rm p})\cdot\db=\omega_{\rm p}^3|\db|^2/3\pi\hbar\epsilon_0 c^3$, where $\GG_0$ is the vacuum's Green's function [i.e. the solution to Eq.~(\ref{gf}) when $\epsilon(\rb,\omega)=1$]. Depending on the geometry and dielectric response of the nanostructure, and on the atom position, $\Gamma'$ can be larger or smaller than $\Gamma_0$. $J'$ accounts for frequency shifts due to other guided and non-guided modes, and is in general spatially dependent. For an atom placed in free-space, $J'$ is the Lamb shift, which renormalizes the atomic resonance frequency, and is considered to be included in the definition of the free-space $\omega_{\rm A}$. The presence of a nanostructure shifts the atomic resonance frequency. We will for simplicity consider this shift identical for every atom and assume its value to be zero.

Once the dynamics of the atomic coherences are solved for, one can reconstruct the field at any point in space. Generalizing Eq.~(6.16) of Ref.~\cite{DKW02} for more than a single atom, the evolution of the bosonic field operator is given by
\begin{align}\label{boson}
\dot{\hat{\textbf{f}}}\,(\rb,\omega)&=-\ii\omega\, \hat{\textbf{f}}(\rb,\omega)\\\nonumber
&+
\frac{\omega^2}{c^2}\sqrt{\frac{1}{\pi\hbar\epsilon_0}\text{Im}\{\epsilon(\rb,\omega)}\}\sum_{j=1}^N \GG^*(\rb,\rb_j,\omega)\cdot \db\,\hge^j,
\end{align}
where the atoms act as sources for the bosonic fields. We can formally integrate this expression and plug it into the equation for the field [Eq.~(\ref{field})]. After some algebra, and performing Markov's approximation, we arrive at the final expression for the field operator, which is simply
\begin{align}\label{fielddef}
\hat{\Eb}^+(\rb)&=\hat{\Eb}_{\rm p}^+(\rb)+\mu_0 \omega^2_{\rm p} \sum_{j=1}^N \GG(\rb,\rb_j,\omega_{\rm p})\cdot\db\,\hge^j.
\end{align}
This expression can be understood as a generalized input-output equation, where the total guided mode field is the sum of the probe, i.e. free, field $\hat{\Eb}_{\rm p}^+(\rb)$ and the field re-scattered by the atoms. The quantum nature of these equations has been treated before when deriving a generalized input-output formalism for unstructured waveguides \cite{CMS15,XF15}.

\section{Transmission and reflection in quasi-1D systems}
\subsection{Atomic coherences in the low saturation regime}
We now explore the behavior of the atoms under a coherent, continuous-wave probe field. In the single-excitation manifold and low saturation (linear) regime ($\braket{\hat{\sigma}_{ee}}=0$), the atoms behave as classical dipoles. Then, the Heisenberg equations for the expectation value of the atomic coherences ($\braket{\hat{\sigma}_{eg}}=\eg$) are linear on the atomic operators, and read
\begin{equation}\label{sigmas}
\dot{\sigma}_{ge}^i=\ii\left(\Delta_{\rm A}+\ii\frac{\Gamma'}{2}\right)\ge^i+\ii\Omega_i
+\ii\sum_{j=1}^N g_{ij}\,\ge^j,
\end{equation}
where $\Omega_i=\db^*\cdot\Eb_{\rm p}^+(\rb_i)/\hbar$ is the guided mode Rabi frequency (with $\Eb_{\rm p}=\braket{\hat{\Eb}_{\rm p}}$), and 
\[g_{ij}=J_{\text{1D}}^{ij}+\ii\Gamma_{\text{1D}}^{ij}/2\,=(\mu_0\omega_{\rm p}^2/\hbar)\,\db^*\cdot\mathbf{G}_{\rm 1D}(\rb_i,\rb_j,\omega_{\rm p})\cdot\db\]
depends only on the Green's function of the guided mode. For long times, the coherences will damp out to a steady state ($\dot{\sigma}_{ge}^i=0$). The solution for the atomic coherences is then
\begin{align}\label{sigmam}
\vec{\sigma}_{ge}=-\mathcal{M}^{-1}\Omegab \;\text{ with  }\;\mathcal{M}=\left(\Delta_{\rm A}+\ii\Gamma'/2\right)\mathbb{1}+\mathfrak{g}.
\end{align}
In the above equation, $\vec{\sigma}_{ge}=(\ge^1,\,\hdots\,,\ge^N)$ and $\Omegab=(\Omega_1,\,\hdots\,,\Omega_N)$ are vectors of $N$ components, and $\mathcal{M}$ is a $N\times N$ matrix that includes the dipole-projected matrix $\mathfrak{g}$ of elements $g_{ij}$. Significantly, the matrix is not Hermitian, as there is radiation loss. However, due to reciprocity, the Green's function matrix is complex symmetric [$\GG^{\rm T}(\rb,\rb',\omega)=\GG(\rb',\rb,\omega)$], and $\mathfrak{g}$ inherits this property if the dipole matrix elements are real, which will be a condition enforced from now on. Complex symmetric matrices can be diagonalized, $\mathfrak{g}\textbf{v}_{\xi}=\lambda_{\xi}\textbf{v}_{\xi}$ with $\xi=1\ldots N$, where $\lambda_{\xi}$ and $\textbf{v}_{\xi}$ are the eigenvalues and eigenvectors of $\mathfrak{g}$, respectively. Since the first term of $\mathcal{M}$ is proportional to the identity, $\mathcal{M}$ and $\mathfrak{g}$ share the same set of eigenvectors.

The eigenmodes represent the spatial profile of the collective atomic excitation, i.e., the dipole amplitude and phase at each atom. However, as the matrix $\mathfrak{g}$ is non-Hermitian, the eigenmodes are not orthonormal in the regular sense, but instead follow different orthogonality and completeness prescriptions, namely $\textbf{v}_{\xi}^T\cdot\textbf{v}_{\xi'}=\delta_{\xi,\xi'}$ and $\sum_{\xi=1}^N \textbf{v}_{\xi}\otimes\textbf{v}_{\xi}^T=\mathbb{1}$, where $T$ indicates transpose instead of the customary conjugate transpose \cite{HJ13}. 
After inserting the completeness relation into Eq.~(\ref{sigmam}), we find that the expected value of the atomic coherences in the steady state in terms of the eigenvalues and eigenvectors of the quasi-1D Green's function is
\begin{align}\label{sigmaeig}
\vec{\sigma}_{ge}=-\sum_{\xi\in\text{mode}}\frac{(\textbf{v}^T_{\xi}\cdot\Omegab)}{(\Delta_{\rm A}+J_{\xi,\rm 1D})+\ii(\Gamma'+\Gamma_{\xi,\rm 1D})/2}\textbf{v}_{\xi},
\end{align}
where $J_{\xi,\rm 1D}=\text{Re}\,\lambda_{\xi}$ and $\Gamma_{\xi,\rm 1D}=2\,\text{Im}\,\lambda_{\xi}$ are the frequency shifts and decay rates corresponding to mode $\xi$, and the sum is performed over mode number from 1 to $N$.  The scalar product in the numerator $\textbf{v}^T_{\xi}\cdot\vec{\Omega}=\sum_{j=1}^N v_{\xi,j}\,\Omega_j$ describes the coupling between the probe field and a particular collective atomic mode. Both the frequency shifts and decay rates, as well as the spatial profile eigenstates of $\mathfrak{g}$, are frequency dependent.

The dynamics of the atoms can be understood in terms of the eigenmodes of $\mathfrak{g}$, where the real and imaginary parts of the eigenvalues correspond to cooperative frequency shifts and decay rates of the collective atomic modes $\{\xi\}$. As the modes are non-normal, the observables cannot be expressed as the sum over all different mode contributions but, instead, any measurable quantity will show signatures of interference between different modes. Although it could be considered a mathematical detail, the fact that the modes of a system are non-normal has deep physical consequences. For instance, non-normal dynamics is responsible of phenomena as different as the Petermann excess-noise factor observed in lasers \cite{Sb1986,Berry04, Berry03} or the transient growth of the shaking of a building after an earthquake \cite{P15}.

\subsection{Transmission and reflection coefficients}
Having previously calculated the linear response of an ensemble of atoms to an input field, we now relate the response to observable outputs, i.e. the reflected and transmitted fields. One can calculate the total field from Eq.~(\ref{fielddef}), by substituting in the solution of Eq.~(\ref{sigmaeig}) for the atomic coherences $\sigma_{ge}$. For a dipole moment directed along $\alpha$ (i.e. $\db=d\hat{\alpha}$), the $\beta-$polarization component of the field reads
\begin{align}\label{fieldeig}
E^+_\beta(\rb)&=E^{+}_{\rm p,\beta}(\rb)-\sum_{\xi=1}^N \frac{\left(\ggm^T_{\alpha\beta}(\rb)\cdot\textbf{v}_{\xi}\right) \left(\textbf{v}_{\xi}^T \cdot\Eb_{\rm p,\alpha}^+\right)}{(\Delta_{\rm A}+J_{\xi,\rm 1D})+\ii(\Gamma'+\Gamma_{\xi,\rm 1D})/2},
\end{align}
where the $j$-component of the electric field vector $\Eb_{\rm p,\alpha}^+$, which reads $\Eb_{\rm p,\alpha,j}^+=\hat{\alpha}\cdot\Eb_{\rm p}^+(\rb_j)$, no longer represents different polarization components, but the dipole-projected field evaluated at the atoms' positions $\rb_j$.  The $j$-component of vector $\ggm_{\alpha\beta} (\rb)$ is $g_{\alpha\beta,j}(\rb)=g_{\alpha\beta}(\rb,\rb_j)=(\mu_0\omega_{\rm p}^2d^2/\hbar)\hat{\alpha}\cdot\GG_{{\rm 1D}} (\rb,\rb_j,\omega_p)\cdot\hat{\beta}$, where $j$ runs over the atom number. In particular, the scalar product $\ggm_{\alpha\beta}^T(\rb)\cdot\textbf{v}_{\xi}=\sum_{i=j}^N\,g_{\beta\alpha,j}(\rb)v_{\xi,j}$ represents how much the mode $\xi$ contributes to the field emitted by the atoms.

\begin{figure*}
\centerline{\includegraphics[width=\textwidth]{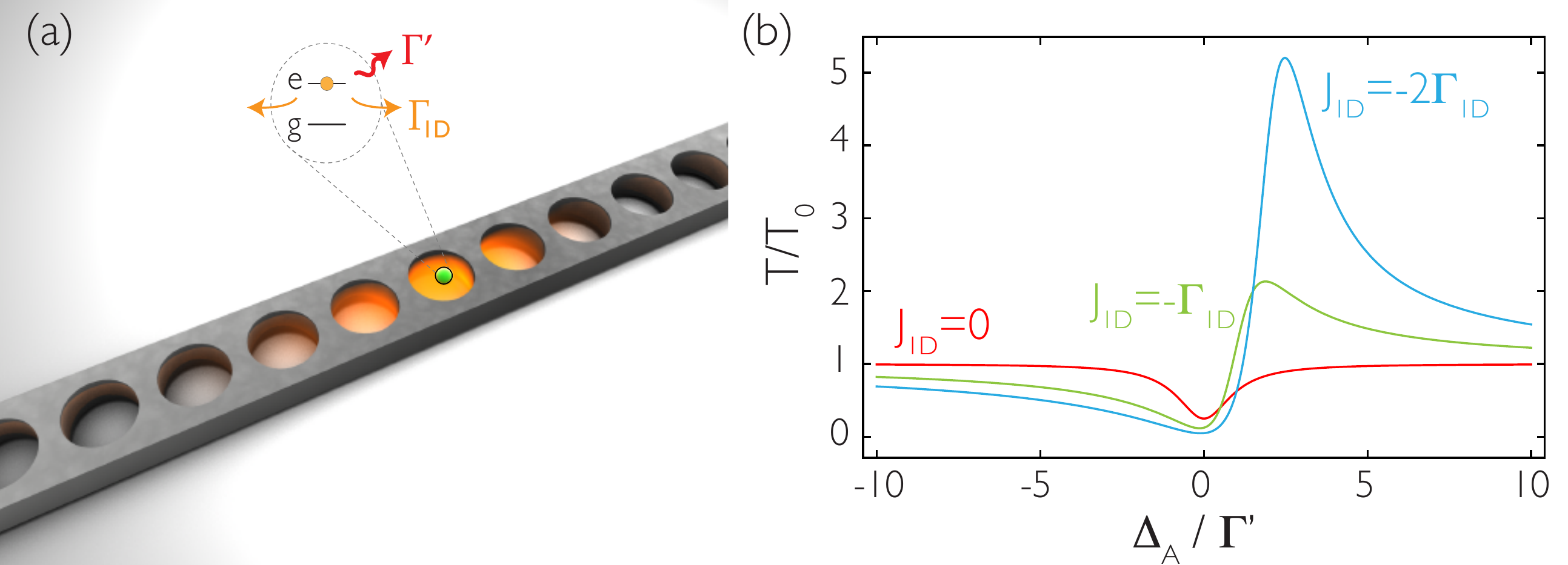}}
\caption{(a) Sketch of an atom interacting with the guided mode of a structured 1D nanostructure. The single-atom decay rate is $\Gamma_{\rm 1D}$, and the decay into non-guided modes is characterized by $\Gamma'$. (b) Normalized transmission spectra ($|t/t_0|^2$) for a single atom for different values of the ratio between the real and imaginary parts of the guided mode Green's function, following Eq.~(\ref{single}). The decay rate into the guided modes is taken to be $\Gamma_{\rm 1D}=\Gamma'$ for all cases.} \label{Fig1}
\end{figure*}

We now assume that the atomic chain and the main axis of the nanostructure are oriented along $\hat{x}$. In order to connect the above expression to the transmission and reflection coefficients, we evaluate the field $E_\beta^+(\rb)$ at the positions $\rb=\rb_{\rm right}$ and $\rb=\rb_{\rm left}$, which are considered to be immediately outside the atomic chain, and only differ in the $x$-component. Following Appendix~\ref{ApA}, the normalized transmission and reflection coefficients are 
\begin{widetext}
\begin{subequations}\label{tandr}
\begin{equation}
\frac{t(\Delta_{\rm A})}{t_0(\Delta_{\rm A})}=1-\frac{1}{g_{\beta\beta}(\rb_{\rm right},\rb_{\rm left})}\sum_{\xi=1}^N \frac{\left(\ggm^T_{\alpha\beta}(\rb_{\rm right})\cdot\textbf{v}_{\xi}\right) \left(\textbf{v}_{\xi}^T \cdot\ggm_{\alpha\beta}(\rb_{\rm left})\right)}{(\Delta_{\rm A}+J_{\xi,\rm 1D})+\ii(\Gamma'+\Gamma_{\xi,\rm 1D})/2},
\end{equation}
\begin{equation}
r(\Delta_{\rm A})=r_0(\Delta_{\rm A})-\frac{1}{g_{\beta\beta}(\rb_{\rm left},\rb_{\rm left})}\sum_{\xi=1}^N \frac{\left(\ggm^T_{\alpha\beta}(\rb_{\rm left})\cdot\textbf{v}_{\xi}\right) \left(\textbf{v}_{\xi}^T \cdot\ggm_{\alpha\beta}(\rb_{\rm left})\right)}{(\Delta_{\rm A}+J_{\xi,\rm 1D})+\ii(\Gamma'+\Gamma_{\xi,\rm 1D})/2},
\end{equation}
\end{subequations}
\end{widetext}
where $t_0(\Delta_{\rm A})$ and $r_0(\Delta_{\rm A})$ are the transmission and reflection coefficients for the 1D photonic structure when no atoms are present. 

\subsection{Simplified expression for the transmission}
For linearly-polarized, transverse guided modes with an approximately uniform transverse field distribution, one can further simplify the expression for the transmission coefficient. We find a product equation that only depends on the eigenvalues of the Green's function matrix $\mathfrak{g}$, and not on their spatial structure (i.e., the eigenfunctions). Following Appendix~\ref{ApB}, we obtain
\begin{align}\nonumber
\frac{t(\Delta_{\rm A})}{t_0(\Delta_{\rm A})}&=\prod_{\xi=1}^N\,\frac{\Delta_{\rm A}+\ii\Gamma'/2}{(\Delta_{\rm A}+J_{\xi,\rm 1D})+\ii(\Gamma'+\Gamma_{\xi,\rm 1D})/2}\\\label{transmission}
&\equiv \prod_{\xi=1}^N\,t_{\xi}(\Delta_{\rm A}).
\end{align}

The total transmission coefficient can thus be written as the product of the transmission coefficients of each of the collective atomic modes. Noticeably, when looking at the transmission spectrum of atoms that interact through the guided mode of a quasi-1D nanostructure, there is a redundancy between the eigenfunctions and eigenvalues, and one is able to obtain an expression that does not depend on the former (i.e., all the relevant information about the geometry is contained in the collective frequency shifts and decay rates). In particular, for a single atom located at $x_j$ with $J^{jj}_{\rm 1D}\equiv J_{\rm 1D}$ and $\Gamma^{jj}_{\rm 1D}\equiv \Gamma_{\rm 1D}$, the eigenvalues are directly proportional to the local Green's function, and
\begin{align}\label{single}
\frac{t(\Delta_{\rm A})}{t_0(\Delta_{\rm A})}=\frac{\Delta_{\rm A}+\ii\Gamma'/2}{(\Delta_{\rm A}+J_{\rm 1D})+\ii(\Gamma'+\Gamma_{\rm 1D})/2}.
\end{align}
The transmittance $T=|t|^2$ can be recast into a Fano-like lineshape \cite{F1936} as
\begin{align}
\frac{T}{T_0}=\frac{(q+\chi)^2}{1+\chi^2}+\left(\frac{\Gamma'}{\Gamma'+\Gamma_{\rm 1D}}\right)^2\frac{1}{1+\chi^2},
\end{align}
where $\chi=2(\Delta_{\rm A}+J_{\rm 1D})/(\Gamma_{\rm 1D}+\Gamma')$ and $q=-2J_{\rm 1D}/(\Gamma_{\rm 1D}+\Gamma')$ is the so-called asymmetry parameter. For $\Gamma'\ll \Gamma_{\rm 1D}$, the second term is negligible and the normalized transmittance is a pure Fano resonance, with $q=-\text{Re}\{\text{G}_{\rm 1D}(\rb_j,\rb_j,\omega_{\rm p})\}/\text{Im}\{\text{G}_{\rm 1D}(\rb_j,\rb_j,\omega_{\rm p})\}$. Fano resonances arise whenever there is interference between two different transport channels. For instance, in a cavity far from resonance, there is interference arising from all the possible optical paths that contribute to the transmission signal due to reflections at the mirrors, whereas in an unstructured waveguide there is no such interference and thus the lineshape is Lorentzian.

For a single atom, there is a clear mapping between the spectrum lineshape and the local 1D Green's function. For a nanostructure with a purely imaginary self Green's function $\GG(x_i,x_i)$ (such as a wave-guide or a cavity at resonance), the spectrum is Lorentzian, and centered around the atomic frequency. However, if the real part is finite, one would observe a frequency shift of the spectrum, which becomes asymmetric. Figure~\ref{Fig1}(b) shows how the normalized transmission spectrum for a single atom becomes more and more asymmetric for higher ratios  $J_{\rm 1D}/\Gamma_{\rm 1D}$. Also, there is an appreciable blueshift of the spectral features.

We would like to remark that the Markov approximation has thus far been employed in our analysis, as every Green's function is considered to be a complex constant over frequency ranges larger than the linewidth of the atoms. However, the expressions for the transmission and reflection coefficients are valid also in the non-Markovian regime. We analyze this issue in more detail in Appendix~\ref{ApC}.

\section{Application to several one-dimensional photonic structures}
In this section, we analyze the transmission spectra of atoms placed along common quasi-1D nanostructures, such as cavities, waveguides, and photonic crystals. 

\subsection{Standing-wave cavities}
To begin with, we want to illustrate the connection between the Green's function formalism and the well-known Jaynes Cummings (JC) model \cite{MS1990,BW08}. For $N$ atoms in a driven cavity of length $L$ and effective area $A$, the JC Hamiltonian, and its corresponding Lindblad operator read 
\begin{subequations}
\begin{align}\nonumber
\mathcal{H}&=-\hbar\Delta_{\rm c}\hat{a}^\dagger\hat{a}-\hbar\Delta_{\rm A}\sum_{i=1}^N \hat{\sigma}_{ee}^i+\hbar\sum_{i=1}^N \mathcal{q}_i\left(\hat{a}^\dagger\hat{\sigma}_{ge}^i+\hat{\sigma}_{eg}^i\hat{a}\right)\\\label{hjc}
&+\hbar\eta \,(\hat{a}+\hat{a}^\dagger),
\end{align}
\begin{align}\nonumber
\mathcal{L}[\hat{\rho}]&=\frac{\Gamma'}{2}\,\sum_{i,j=1}^N\left(2\hat{\sigma}_{ge}^i\hat{\rho}\hat{\sigma}_{eg}^j-\hat{\sigma}_{eg}^i\hat{\sigma}_{ge}^j\hat{\rho}-\hat{\rho}\hat{\sigma}_{eg}^i\hat{\sigma}_{ge}^j\right)\\\label{ljc}
&+\frac{\kappa_c}{2}\,\left(2\hat{a}\hat{\rho}\hat{a}^\dagger-\hat{a}^\dagger\hat{a}\hat{\rho}-\hat{\rho}\hat{a}^\dagger\hat{a}\right),
\end{align}
\end{subequations}
where $\hat{a}$ is the cavity-field annihilation operator, $\hat{\rho}$ is the density matrix for the atoms and the cavity field, $\eta$ is a frequency that represents the amplitude of the classical driving field, $\Delta_{\rm c}=\omega_{\rm p}-\omega_{\rm c}$ is the detuning between the driving (probe) and the cavity fields, and $\kappa_c$ is the cavity-field decay. The atom cavity coupling is $\mathcal{q}_i=\mathcal{q}\cos(k_{\rm c} x_i)$, where $\mathcal{q}=d\sqrt{\omega_{\rm c}/(\hbar\epsilon_0 LA)}$ is modulated by a function that depends on the atoms' positions and the cavity wave-vector $k_{\rm c}$. The Heisenberg equations of motion for the field and atomic operators are
\begin{subequations}
\begin{align}
&\dot{\hat{a}}=\left(\ii\Delta_{\rm c}-\frac{\kappa_{\rm c}}{2}\right)\hat{a}-\ii\sum_{i=1}^N \mathcal{q}_i\hge^i-\ii\eta,\\
&\dot{\hat{\sigma}}_{ge}^i=\left(\ii\Delta_{\rm A}-\frac{\Gamma'}{2}\right)\hge^i+\ii\mathcal{q}_i\left(\hat{\sigma}_{ee}^i-\hat{\sigma}_{gg}^i\right)\hat{a}.
\end{align}
\end{subequations}
When $ \Gamma'\ll\kappa_{\rm c}$ and $\mathcal{q}<\text{min}\{\Delta_{\rm c},\kappa_{\rm c}\}$, the cavity field can be adiabatically eliminated, and the field operator re-expressed in terms of the atomic ones, i.e.,
\begin{align}\nonumber
\dot{\hat{a}}=0\quad\rightarrow\quad \hat{a}=\frac{1}{\left(\Delta_{\rm c}+\ii\frac{\kappa_{\rm c}}{2}\right)}\left(\eta+\sum_{i=1}^N \mathcal{q}_i\hge^i\right).
\end{align}
Introducing this expression back into the equation for the atomic operator, one can deduce a master equation for the atomic density matrix $\hat{\rho}_{\rm A}$. The new Hamiltonian and Lindblad operators read just as those of Eqs.~(\ref{ham}) and (\ref{lind}), but for a classical driving field, and with spin exchange and decay rates into the cavity mode given by \cite{GP97}
\begin{subequations}
\begin{align}
J^{ij}_{\rm 1D}&=-\frac{\mathcal{q}^2\Delta_{\rm c}}{(\Delta_{\rm c}^2+\kappa_{\rm c}^2/4)}\cos(k_{\rm c} x_i)\cos(k_{\rm c} x_j),\\
\Gamma^{ij}_{\rm 1D}&=\frac{\mathcal{q}^2\kappa_{\rm c}}{(\Delta_{\rm c}^2+\kappa_{\rm c}^2/4)}\cos(k_{\rm c} x_i)\cos(k_{\rm c} x_j).
\end{align}
\end{subequations}
It can thus be seen that the Markovian approximation to arrive at these equations is equivalent to the absence of strong coupling effects within the JC model.

The last step for connecting this simple model with our formalism is to calculate the Green's function of a cavity and confirm that $J^{ij}_{\rm 1D}$ and $\Gamma^{ij}_{\rm 1D}$ are precisely those obtained within the JC framework. The Green's function of a quasi-1D cavity formed by partially transmitting mirrors of reflection coefficient $r$ (chosen to be real) is \cite{HGM13}
\begin{align}\nonumber
&G_{\rm 1D}(x_i,x_j,\omega_{\rm p})\simeq\frac{\ii c^2}{2v_g\omega_p A (1-r^2 e^{2\ii k_{\rm p} L})}\left[e^{\ii k_{\rm p} |x_i-x_j|}\right.\\
&\left.+re^{\ii k_{\rm p}(L+x_i+x_j)}+re^{\ii k_{\rm p}[L-(x_i+x_j)]}+r^2e^{\ii k_{\rm p}(2L-|x_i-x_j|)}\right],
\end{align}
where $v_{\rm g}$ is the group velocity. For high-Q standing-wave cavities, i.e. with $r\simeq 1$, and choosing $v_{\rm g}=c$, the Green's function can be approximated as
\begin{align}\nonumber
G_{\rm 1D}(x_i,x_j,\omega_{\rm p})\simeq &\left(\frac{2\ii c}{\omega_p A}\right)\frac{1}{1-r^2e^{2\ii k_{\rm p} L}}\\
&\times\cos(k_{\rm p} x_i) \cos(k_{\rm p} x_j).
\end{align}
The cavity is resonant at a frequency $\omega_{\rm c}$ with corresponding wave-vector $k_{\rm c}$, chosen to be such that $k_{\rm c} L=2\pi m$, with $m$ being an integer. Close to resonance, one can write $k_{\rm p}=k_{\rm c}+\delta k$, and assume that $\delta k L\ll1$. Then  $1-r^2e^{2\ii k_{\rm p} L}\simeq 1-r^2-2\ii r^2\delta k L$, and the Green's function is simply
\begin{align}
G_{\rm 1D}(x_i,x_j,\omega_{\rm p})
\simeq -\left(\frac{c^2}{\omega_{\rm p} L A}\right)\frac{\cos(k_{\rm c} x_i) \cos(k_{\rm c} x_j)}{\Delta_c+\ii\kappa_c/2},
\end{align}
where $\kappa_c=(1-r^2) c/L$ is the cavity linewidth. Therefore, the atoms' spin-exchange and decay rates are given by 
\begin{subequations}
\begin{align}\nonumber
J^{ij}_{\rm 1D}&=\frac{\mu_0\omega_{\rm p}^2 d^2}{\hbar}\,\text{Re}\,G_{\rm 1D}(x_i,x_j,\omega_{\rm p})=-\mathcal{q}_i\mathcal{q}_j\frac{\Delta_{\rm c}}{(\Delta_{\rm c}^2+\kappa_{\rm c}^2/4)},
\end{align}
\begin{align}\nonumber
\Gamma^{ij}_{\rm 1D}&=\frac{2\mu_0\omega_{\rm p}^2 d^2}{\hbar}\,\text{Im}\,G_{\rm 1D}(x_i,x_j,\omega_{\rm p})=\mathcal{q}_i\mathcal{q}_j\frac{\kappa_c}{(\Delta_{\rm c}^2+\kappa_{\rm c}^2/4)},
\end{align}
\end{subequations}
which is precisely what is obtained within the Jaynes Cummings model. 
\begin{figure*}
\centerline{\includegraphics[width=\textwidth]{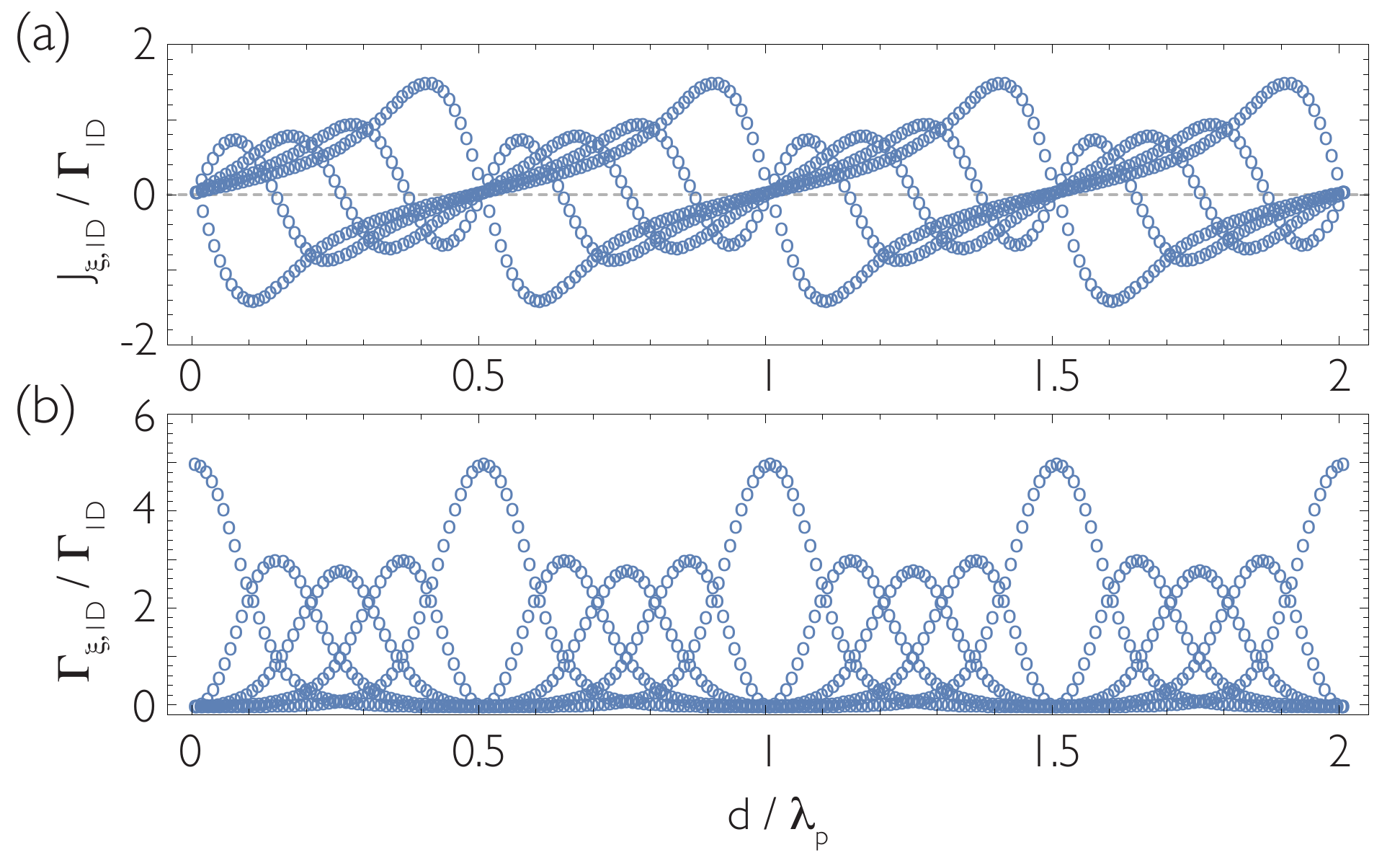}}
\caption{(a) Frequency shifts and (b) decay rates of the collective modes of a regular chain of 5 atoms placed along a waveguide normalized to the single-atom decay rate into the guided mode $\Gamma_{\rm 1D}$, as a function of the distance $d$ between the atoms in units of the probe wavelength.} \label{Fig2}
\end{figure*}
Let's now look at the transmission spectrum of $N$ atoms in a cavity. As we have just demonstrated, coefficients of the dipole-projected Green's function matrix $\mathfrak{g}$ read
\begin{equation}\label{hh}
g_{ij}=g(\omega_{\rm p})\cos(k_{\rm c} x_i) \cos(k_{\rm c} x_j),
\end{equation}
where $g(\omega_{\rm p})=J^{\rm max}_{\rm 1D}+\ii\Gamma^{\rm max}_{\rm 1D}/2$, where $J^{\rm max}_{\rm 1D}$ and $\Gamma^{\rm max}_{\rm 1D}$ are the spin-exchange and decay rates at the antinode of the cavity field. Depending on the detuning between the probe field and the cavity resonance, $g(\omega_{\rm p})$ can be purely imaginary, yielding dissipative atom-atom interactions, or can have both real and imaginary parts, resulting in both dissipative and dispersive couplings.

The matrix $\mathfrak{g}$ is separable (has rank one) as it can be written as the tensor product of just one vector by itself. The matrix has one eigenstate describing a superposition of atomic coherences that couples to the cavity (a "bright mode"), with eigenvalue $\lambda_{\rm B}=\sum_{i=1}^Ng^{ii}= \left(J^{\rm max}_{\text{1D}}+\ii\Gamma^{\rm max}_{\text{1D}}/2\right)\sum_{i=1}^N \cos^2(k_{\rm c}x_i)$. This atomic collective excitation follows spatially the mode profile of the cavity, i.e. $\ge^i\propto\cos(k_{\rm c} x_i)$. The matrix $\mathfrak{g}$ has also $N-1$ decoupled ("dark") modes of eigenvalue 0. Because these dark modes have a zero decay rate into the cavity mode, it is also impossible to excite them employing the cavity field. The optical response is thus entirely controlled by the bright mode, and the transmission is simply
\begin{align}
\frac{t(\Delta_{\rm A})}{t_0(\Delta_{\rm A})}=\frac{\Delta_{\rm A}+\ii\Gamma'/2}{(\Delta_{\rm A}+\sum_{i=1}^N  J^{ii}_{\rm 1D})+\ii(\Gamma'+\sum_{i=1}^N  \Gamma^{ii}_{\rm 1D})/2}.
\end{align}
Remarkably, this expression is valid no matter the separation between the atoms or whether they form an ordered or disordered chain. The transmission spectrum corresponds to that of a `super-atom', where the decay rates and the frequency shifts are enhanced (N-fold if all the diagonal components of $\mathfrak{g}$ are equal) compared to those of a single atom. This result replicates the well-known expressions for conventional cavity QED. 

\subsection{Unstructured Waveguides}
Another paradigm that has been investigated frequently is that of "waveguide QED" \cite{SF05}. The simple model of such a system consists of a single guided mode with translational invariance, and where the dispersion relation is well-approximated as linear around the atomic resonance frequency. In a 1D translationally invariant system, a source simply emits a plane wave whose phase at the detection point is proportional to the distance of separation. Therefore, the elements of the Green's function matrix $\mathfrak{g}$ depend on the distance between the atoms, and read  
\begin{equation}\label{gwg}
g_{ij}=\ii\frac{\Gamma_{\rm 1D}}{2}e^{\ii k_p |x_i-x_j|}.
\end{equation}
Remarkably, the self Green's function in a waveguide is purely imaginary. The coherent interactions between atom $i$ and atom $j$ are dictated by the Hamiltonian [given by Eq.~(\ref{ham})], and are proportional to $\text{Re}\{g_{ij}\}=-(\Gamma_{\rm 1D}/2)\sin k_{\rm p}|x_i-x_j|$, whereas the dissipation is given by the Lindblad operator [given by Eq.~(\ref{lind})], which is proportional to $\text{Im}\{g_{ij}\}=(\Gamma_{\rm 1D}/2)\cos k_{\rm p}(x_i-x_j)$ \cite{GMM11,CJG12}. It is thus clear that by carefully tuning the distance between the emitters, one can engineer fully dissipative interactions. If the atoms form a regular chain and are spaced by a distance $d$ such that $k_pd=n\pi$, where $n$ is an integer number, the matrix $\mathfrak{g}$ has only one non-zero eigenvalue $\lambda_{\rm B}=\ii N\Gamma_{\rm 1D}/2$ associated with the bright atomic mode. This situation is analogous to the case of atoms interacting in an on-resonance cavity. Therefore, there will not be any collective frequency shift, and the lineshape will be a Lorentzian of width $N\Gamma_{\rm 1D}+\Gamma'$. For $n$ even, the phases of the dipole moments of the atoms are all identical, whereas for odd $n$ the dipole moments of adjacent atoms are $\pi$ out of phase.

\begin{figure}
\centerline{\includegraphics[width=\linewidth]{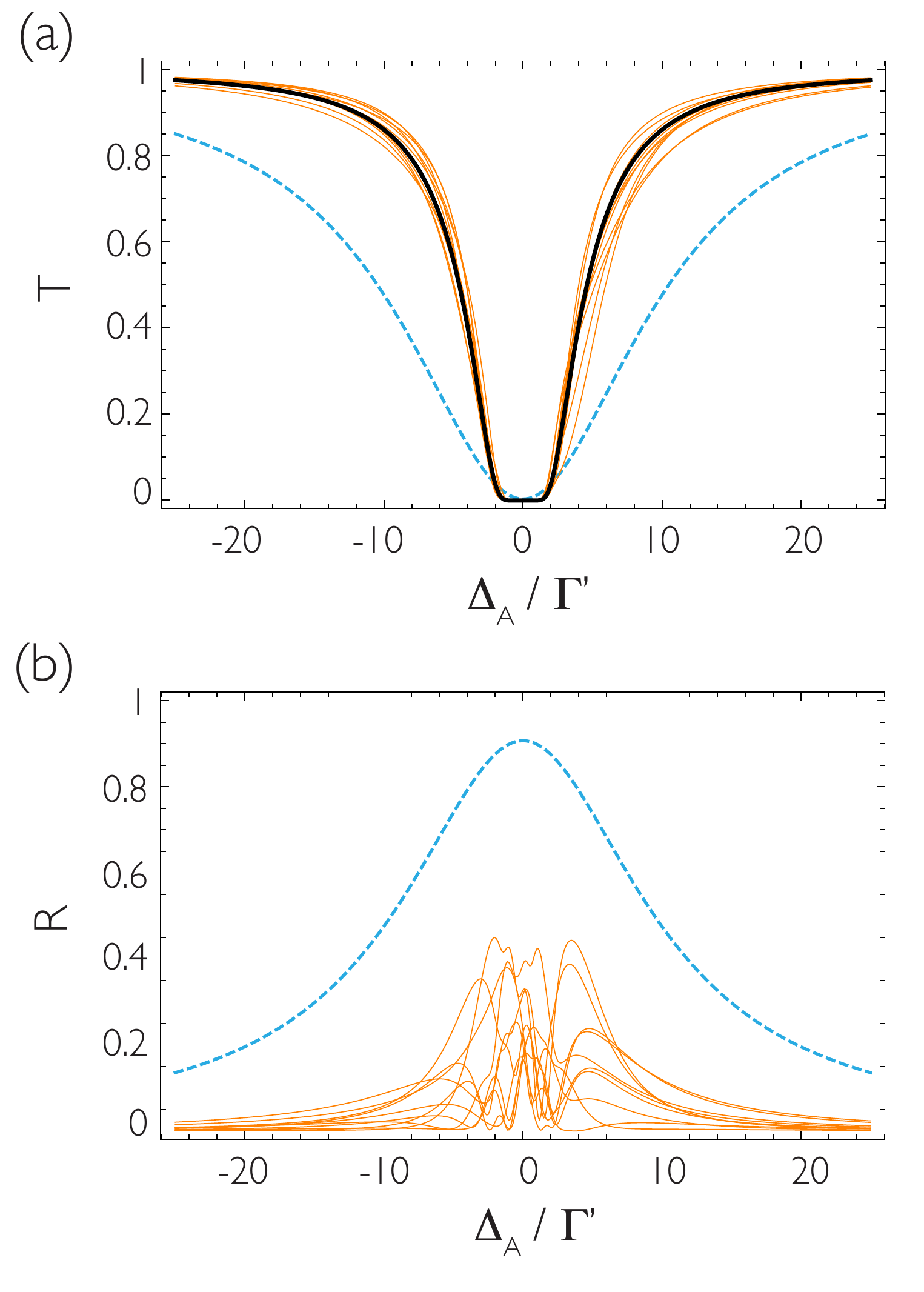}}
\caption{(a) Transmission spectra for 20 atoms interacting through the guided modes of an unstructured waveguide. The dashed blue line represents a regular separation between the atoms of $d=\lambda_{\rm p}/2$. The orange curves show 10 different spectra obtained by randomly placing the atoms along the nanostructure. The thick black curve represents the "non-interacting" case of Eq.~(\ref{tindep}).
(b) Reflection spectra for the same situations as in (a). We have chosen $\Gamma_{\rm 1D}=\Gamma'$, $T_0=1$, and $R_0=0$.} \label{Fig3}
\end{figure}

For a regular chain with lattice constant different from $k_{\rm p}d=n\pi$, or for atoms placed randomly along the waveguide, the coefficients of matrix $\mathfrak{g}$ have both a real and imaginary part, and, to the best of our knowledge, there is no analytic expression for the eigenvalues of $\mathfrak{g}$. Figure~\ref{Fig2} shows the frequency shifts and decay rates of the collective modes of a $N=5$ atom chain as a function of the separation between the atoms. For separations where $k_{\rm p}d=n\pi$, the real part of the Green's function is zero and the imaginary part of all modes but one goes to zero, whereas for other spacings one generically gets a zoo of coherent and dissipative couplings of comparable strength. This occurs because the real and imaginary parts of $g_{ij}$ are generically of similar magnitude. Figure~\ref{Fig3} shows the transmission and reflection spectra for $N=20$ atoms separated by $k_{\rm p}d=\pi$ (blue dashed curve), and for several random realizations where each atomic position is chosen randomly from a distribution $k_{\rm p} x_{i} \in [0,2\pi]$ (orange curves). The thick black line in Fig.~\ref{Fig3}(a) represents the non-interacting case, which is obtained by setting the non-diagonal terms of $\mathfrak{g}$ to zero, yielding a transmission spectrum
\begin{align}\label{tindep}
\frac{t(\Delta_{\rm A})}{t_0(\Delta_{\rm A})}=\left(\frac{\Delta_{\rm A}+\ii\Gamma'/2}{\Delta_{\rm A}+\ii(\Gamma'+\Gamma_{\rm 1D})/2}\right)^N,
\end{align}
where the transmission coefficient is a product of the transmission coefficient of each single atom, and the frequency shifts and decay rates are not collective quantities but, instead, single-atom parameters. 

\begin{figure*}
\centerline{\includegraphics[width=\textwidth]{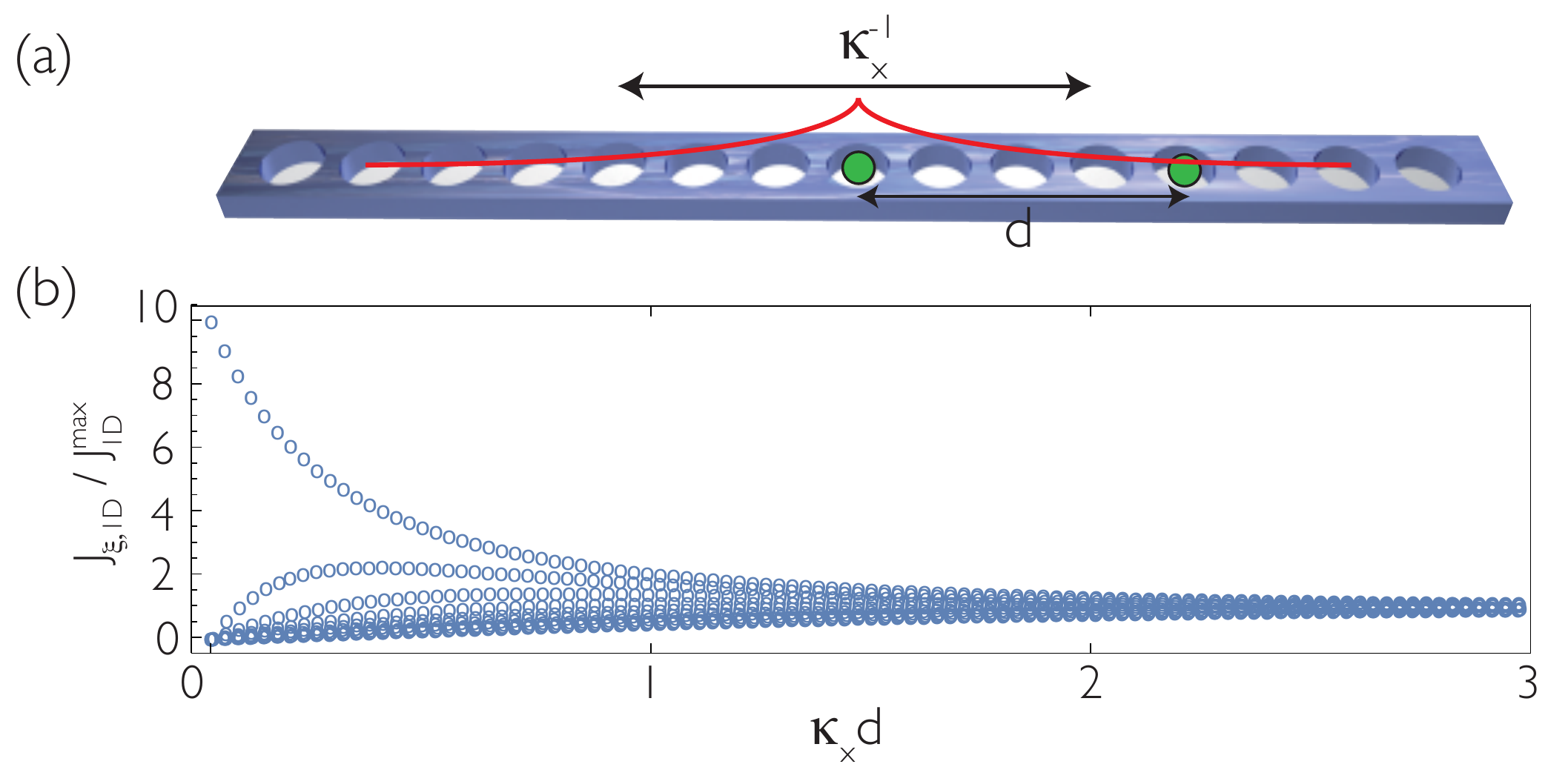}}
\caption{(a) When the atomic resonance frequency lies within the band-gap region of a photonic crystal, an excited atom becomes dressed by a photonic ``cloud'' of tunable size.  A second atom sitting at a distance $d$ interacts with this cloud, giving rise to an effective atomic interaction of spatial range $\kappa_x^{-1}$. (b) Collective frequency shifts of the modes of a regular chain of $N$=10 atoms in the bandgap of an infinite photonic crystal as a function of $\kappa_xd$. The atoms are placed at even antinodes of the Bloch modes.} \label{Fig4}
\end{figure*}

Figure~\ref{Fig3}(a) also shows that, for random filling, although the atoms interact with each other  ($\mathfrak{g}_{ij\neq i}\neq0$), the transmission spectra follow closely that of a non-interacting system, for which all the off-diagonal elements are zero ($\mathfrak{g}_{ij\neq i}=0$), and the eigenvalues of matrix $\mathfrak{g}$ are proportional to the self Green's functions [$\GG(x_i,x_i)$] at the atoms' positions. In this case, the behavior of the emitters cannot be understood in terms of the `super-atom' picture, as the transmission spectrum of the system is significantly different from a Lorentzian. In particular, for the non-interacting scenario, one can recast Eq.~(\ref{tindep}) into an exponential, and the transmittance recovers the well-known form of a Beer-Lambert law, reading
\begin{align}\nonumber
\frac{T(\Delta_{\rm A})}{T_0(\Delta_{\rm A})}&=\text{exp}\left[-N \ln\frac{\Delta_{\rm A}^2+(\Gamma'+\Gamma_{\rm 1D})^2/4}{\Delta_{\rm A}^2+\Gamma'^2/4}\right]\\
&\simeq \text{exp}\left[-\frac{\text{OD}}{1+(2\Delta_{\rm A}/\Gamma')^2}\right],
\end{align}
where $\text{OD}\equiv 2N\Gamma_{\rm 1D}/\Gamma'$ is the optical depth and the last equality holds for $\Gamma_{\rm 1D}\ll\Gamma'$. This is exactly the same behavior that an atomic ensemble in free space would exhibit. This occurs only for non-negligible $\Gamma'$, which suppresses multiple reflections. Otherwise one would see huge fluctuations associated with Anderson localization in the spectra. 

The reflectance spectrum, on the other hand, is more complex and carries more information than the transmittance, as shown in Fig.~\ref{Fig3}(b). In contrast to the case of the transmission coefficient, the reflection does not admit a simple formula in terms of the eigenvalues of the system. This is only possible when the Green's function is separable, namely, when the distance between the atoms is $d=n\lambda_{p}/2$.

\subsection{Photonic crystal bandgaps}
The band-gap region of a photonic crystal waveguide (PCW) is a very appealing scenario to explore coherent atom-atom interactions, as light cannot propagate, and atoms interact with each other through evanescent fields \cite{DHH15}, as depicted in Fig.~\ref{Fig4}(a). Band-gap interactions facilitate long range dispersive dynamics, enabling the engineering of many-body states for atoms and photons with very low dissipation.

For a photonic crystal waveguide of lattice constant $a$ the elements of matrix $\mathfrak{g}$ are well approximated by
\begin{equation}
g_{ij}=J^{\rm max}_{\rm 1D}\cos(\pi x_i/a) \cos(\pi x_j/a) e^{-\kappa_x|x_i-x_j|},
\end{equation}
where the cosine terms account for the spatial profile of the Bloch modes, $J^{\rm max}_{\rm 1D}$ is the value of the spin-exchange rate for an atom located at the maximum of the Bloch mode, and $\kappa_x^{-1}$ is the finite range of interaction due to the evanescent decay of the guided mode field in the bandgap, which is controlled by detuning the band-edge frequency from the atomic resonance. The value of $J^{\rm max}_{\rm 1D}$ is also determined by this detuning. It should be noted that in this idealized picture, $g_{ij}$ is purely real, indicating the absence of collective emission into the PCW. This is naturally expected, due to the absence of guided modes at the atomic frequency. In practice, residual decay might still exist to the extent that the mediating photon has a decay channel. This could be either due to the finite length of the PCW, which can cause the photon to leak out the ends and is suppressed when $\kappa_{x}L \gg 1$, or through scattering and absorption losses of the PCW. Given that these photonic decay processes can be made small, for conceptual simplicity here we treat the idealized case.

For a chain of periodically spaced atoms placed in even antinodes of the Bloch modes, the dipole-projected Green's function matrix reads
\begin{align}\label{Gchi}
\mathfrak{g}=J^{\rm max}_{\rm 1D}\begin{pmatrix}
 1 & \chi & \chi^2 &\cdots & \chi^{N-1} \\
 \chi & 1 & \chi &\cdots & \chi^{N-2}\\
 \vdots  & \vdots & \vdots  &\ddots  & \vdots\\
  \chi^{N-1}  & \chi^{N-2}  & \chi^{N-3}   &\cdots & 1
 \end{pmatrix},
\end{align}
where we have defined $\chi\equiv e^{-\kappa_xd}$, with $d$ being the distance between nearest-neighbor atoms. The matrix $\mathfrak{g}$ is a real symmetric Toeplitz matrix (or bisymmetric matrix). Neglecting higher order contributions besides first-neighbor, an approximation valid for $\kappa_xd\gg 1$, $\mathfrak{g}$ becomes a tridiagonal Toeplitz matrix whose eigenvalues and eigenvectors are \cite{NPR13}:
\begin{subequations}
\begin{equation}
\lambda_{\xi}\equiv J_{{\rm 1D},\xi}=J^{\rm max}_{\rm 1D}+2J^{\rm max}_{\rm 1D}e^{-\kappa_xd}\cos\left(\frac{\xi\pi}{N+1}\right),
\end{equation}
\begin{equation}
v_{\xi,j}=\sqrt{\frac{2}{N+1}}\sin\left(\frac{\xi j \pi}{N+1}\right).
\end{equation}
\end{subequations}

In this simple tight binding model, the frequency shifts of the collective atomic modes are distributed around $J^{\rm max}_{\rm 1D}$ with a frequency spread controlled by $\kappa_{x}$ (i.e., for larger $\kappa_x$, the modes are closer in frequency). This can be observed in Fig.~\ref{Fig4}(b), which shows how the collective frequency shifts coalesce towards $J^{\rm max}_{\rm 1D}$ for large $\kappa_x d$. However, if the interaction length is very large compared to the distance between the atoms, the approximation of neglecting higher order neighbors falls apart, and the eigenvalues start to show a different behavior. Eventually, when the interaction length becomes infinite (or much larger than the length of the atomic distribution), there is only one bright mode, of eigenvalue $\lambda_{\rm B}=NJ^{\rm max}_{\rm 1D}$. This is analogous to the cavity case, where the interaction range is also infinite, except now the eigenvalue is purely real. The band-edge of a photonic crystal is thus a cross-over region in which the single bright mode approximation holds and then transitions to another regime where it breaks down, as the guided mode becomes evanescent and decays substantially within the length of the PCW. Importantly, the bandgap of a photonic crystal provides a tunable interaction range, a feature which is unique to this kind of nanostructure, and makes PCWs remarkably different reservoirs from either cavities or unstructured waveguides.

In the final section, we present some predictions for the transmission spectrum of two atoms coupled to a PCW for $\Gamma_{\rm 1D}$ and $J_{\rm 1D}$ values that can be achieved experimentally in the coming years. We hope that the foreseen large coherent couplings between the atoms combined with low dissipation through the guided mode help to stimulate a new generation of experiments that go beyond the current state of the art.

\section{Electromagnetically induced transparency}
\begin{figure}
\centerline{\includegraphics[width=\linewidth]{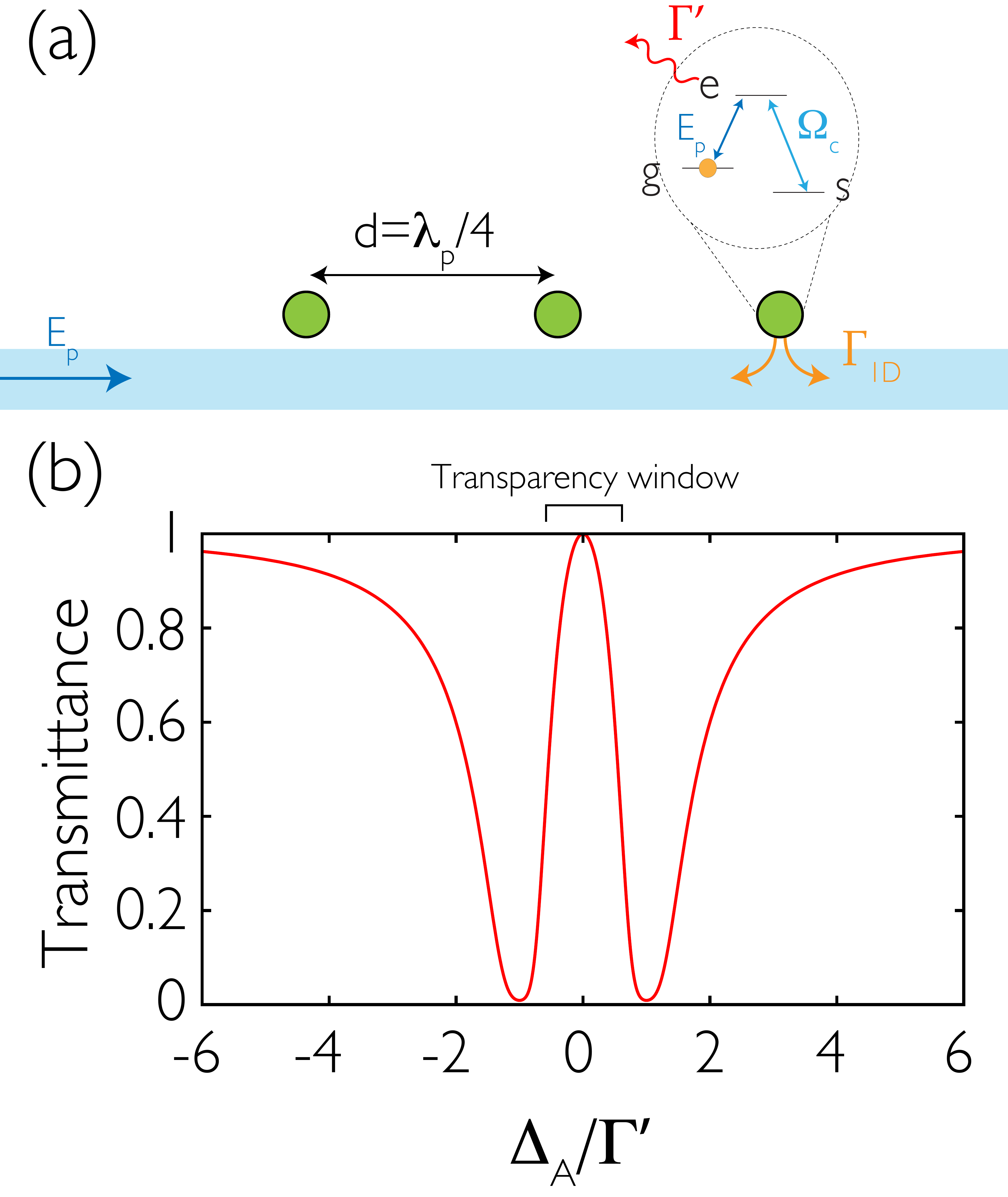}}
\caption{(a) Schematic of electromagnetically-induced transparency. The $\ket{g}$ to $\ket{e}$ transition is coupled to the guided mode of an unstructured waveguide, and the $\ket{s}$ to $\ket{e}$ transition is driven by an external, classical control field of Rabi frequency $\Omega_\text{c}$. (b) Representative transmittance spectrum ($|t_\text{EIT}/t_0|^2$), displaying the characteristic transparency window. The number of atoms is $N=5$, their separation is $d=\lambda_\text{p}/4$, the control field intensity is $\Omega_c=\Gamma'$, the guided mode decay rate is $\ga=0.5 \Gamma'$, and we have set $t_0=1$.} \label{FigEIT}
\end{figure}

Throughout this work, we have dealt with two-level atoms, but our formalism is not restricted to these kind of emitters and simple expressions for light propagation can also be derived in the case of multilevel atoms. In this section we apply the derived expression for the normalized transmission in terms of the collective energy shifts and decay rates [i.e. Eq.~(\ref{transmission})] to one of the paradigmatic problems within quantum optics: electromagnetically induced transparency (EIT) \cite{HFI1990,FL00,FIM05}.

We consider a chain of $\Lambda$-atoms whose $\ket{e}$ to $\ket{g}$ transition is coupled to the guided mode of a quasi-1D nanostructure, as shown in Fig.~\ref{FigEIT}(a). The excited state can also decay to a metastable state $\ket{s}$ of frequency $\omega_s$. This transition, whose polarization is chosen to be orthogonal to the guided mode, is accessed via an external, classical, and uniform control field of Rabi frequency $\Omega_c$. Under the action of the control field, the guided mode photons promote the atoms from the ground state to the so-called dark state, a superposition of both $\ket{g}$ and $\ket{s}$. The creation of these dark states has two main consequences: first, it prevents photon loss due to the the long lifetime of the $\ket{s}$ state; second, the group velocity is greatly reduced as the probe field is converted into a polaritonic excitation made of spins and photons moving at a diminished speed.
 
In order to describe EIT, we have to add another term to the Hamiltonian of Eq.~(\ref{ham}). The new term accounts for the presence of the $\ket{s}$ state and the driving of the $\ket{e}$ to $\ket{s}$ transition, and reads $\mathcal{H}_{\rm c}=-\hbar\sum^N_{i=1}\Delta_{\rm s}\sigma_{ss}^i+\Omega_c\left(\sigma_{es}^i+\sigma_{se}^i\right)$, where $\Delta_{\rm s}=\omega_{\rm p}-\omega_c-\omega_{\rm s}$, and $\omega_c$ is the control field frequency. For the sake of simplicity, we take $\Omega_c$ to be real. In the single-excitation manifold, and for low saturation ($\sigma_{gg}\simeq 1$, and $\sigma_{ss}\simeq\sigma_{ee}\simeq \sigma_{es}\simeq 0$), the equations of motion for the atomic coherences are given by
\begin{subequations}
\begin{align}
&\dot{\sigma}_{ge}^i=\ii\Delta_{\rm A}\ge^i+\ii\Omega_i+\ii\Omega_c\sigma_{gs}^i
+\ii\sum_{j=1}^N g_{ij}\,\ge^j,\\
&\dot{\sigma}_{gs}^i=\ii\Delta_{\rm s}\sigma_{gs}^i+\ii\Omega_c\sigma_{ge}^i,
\end{align}
\end{subequations}
where $\Delta_{\rm A}=\omega_{\rm p}-\omega_{\rm A}$ is the detuning between the guided mode probe field and the atomic $\ket{e}-\ket{g}$ transition. Following the steps described in Section II, we solve these equations in the steady state. Then,  $\sigma_{gs}^i=-(\Omega_c/\Delta_{\rm s})\,\sigma_{ge}^i$, and 
\begin{align}
\vec{\sigma}_{ge}=-\Delta_{\rm s}\sum_{\xi\in\text{mode}}\frac{(\textbf{v}^T_{\xi}\cdot\Omegab)}{\Delta_{\rm s}[\Delta_{\rm A}+\ii\Gamma'/2+\lambda_\xi]-|\Omega_c|^2}\textbf{v}_{\xi}.
\end{align}
Since this equation has the same mathematical structure of Eq.~\eqref{sigmaeig}, one can proceed as before and find the product expression for the normalized transmission, which reads 
\begin{align}\label{teit}
\frac{t_{\rm EIT}}{t_0}=\prod_{\xi=1}^N\,\frac{\Delta_{\rm A}(\Delta_{\rm A}+\ii\Gamma'/2)-\Omega_c^2}{\Delta_{\rm A}[\Delta_{\rm A}+\ii\Gamma'/2+\lambda_\xi]-\Omega_c^2}=\prod_{\xi=1}^N t_{\xi},
\end{align}
where we have chosen  $\Delta_{\rm s}=\Delta_{\rm A}$. When the probe field is on resonance with the $\ket{e}-\ket{g}$ transition, the two-photon resonance condition ($\Delta_{\rm s}=0$) is met, guaranteeing maximal transparency precisely when $\Delta_\text{A}=0$. Figure~\ref{FigEIT}(b) depicts the transmittance spectrum for a chain of $N=5$ atoms separated by a distance $d=\lambda_\text{p}/4$ and coupled to an unstructured waveguide. Close to resonance, the probe field is fully transmitted and the spectrum displays the characteristic transparency window. The limiting result of Eq.~\eqref{teit} when applied to one atom agrees with what has been previously found \cite{SW10}.

From the above equation it is easy to find a general expression for the effective wave-vector of the polaritonic excitation. After a propagation distance spanning $N$ atoms, the transmission becomes $t_{\rm EIT}/t_0=e^{\ii k_{\rm eff} Nd}$, where $k_{\rm eff}$ is a complex number that accounts for both dispersion and absorption. Up to third order in the atom-probe detuning $\Delta_{\rm A}$, we find
\begin{align}\nonumber
k_{\rm eff}&=-\frac{\ii}{Nd}\sum_{\xi=1}^N \lambda_\xi\left\{
\frac{\Delta_{\rm A}}{\Omega_c^2}+\frac{\Delta_{\rm A}^2}{2\Omega_c^4}(\lambda_\xi+\ii\Gamma')\right.\\\label{keff}
&\left.+\frac{\Delta_{\rm A}^3}{12\Omega_c^6}\left[12\Omega_c^2-3\Gamma'^2+6\ii\Gamma' \lambda_\xi+4\lambda_\xi^2\right]\right\}.
\end{align}
The term quadratic in the detuning accounts for absorption, whereas the cubic one describes group velocity dispersion \cite{CSH11}.

The above expression is valid for any linear and isotropic quasi 1D-nanostructure. We would like now to focus on the case of an unstructured waveguide. As we have shown in the previous section, if the atoms are arranged in the mirror configuration (i.e. $k_{\rm p}d=n\pi$, with $n\in \mathbb{N}$) there is only a single eigenstate, of eigenvalue $\lambda_\xi=\ii N\ga/2$, and thus the effective wave-vector is
\begin{align}
k_{\rm eff}&=\frac{\ga\Delta_{\rm A}}{2d\Omega_c^2}+\ii\frac{\ga\Delta_{\rm A}^2}{8d\Omega_c^4}(2\Gamma'+N\ga)\\\nonumber
&+\frac{\ga\Delta_{\rm A}^3}{24d\Omega_c^6}\left[12\Omega_c^2-3N\ga\Gamma'-N^2\ga^2-3\Gamma'^2\right].
\end{align}
For any other configuration, there are $N$ eigenstates, and the calculation of $k_{\rm eff}$ is not as simple. However, realizing that $\sum_{\xi} \lambda_\xi^\beta=\text{Tr} \,\{\mathfrak{g}^\beta\}$, it is easy to evaluate Eq.~(\ref{keff}) for any inter-atomic separation. In particular, when $k_{\rm p}d=n\pi/2$, with $n$ being an odd integer, we find another closed-form solution, namely
\begin{widetext}
\begin{subequations}
\begin{align}
k_{\rm eff}&=\frac{\ga\Delta_{\rm A}}{2d\Omega_c^2}+\ii\frac{\ga\Gamma'\Delta_{\rm A}^2}{4d\Omega_c^4}+\frac{\ga\Delta_{\rm A}^3}{24 d\Omega_c^6}[12\Omega_c^2+2\ga^2-3\Gamma'^2]\quad \text{if $N$ is even},
\end{align}
\begin{align}
k_{\rm eff}&=\frac{\ga\Delta_{\rm A}}{2d\Omega_c^2}+\ii\frac{\ga\Delta_{\rm A}^2}{8d\Omega_c^4}(2\Gamma'+\ga/N)+\frac{\ga\Delta_{\rm A}^3}{24 d\Omega_c^6}[12\Omega_c^2-\ga^2-3\Gamma'^2-3\ga\Gamma'/N]\quad \text{if $N$ is odd}.
\end{align}
\end{subequations}
\end{widetext}
Therefore, for a chain of atoms in a waveguide, the effective polaritonic wavevector scales very differently with the number of atoms depending on the distance between them. For some specific separations, even the parity of the number of emitters modifies the dispersion relation. However, the group velocity at zero detuning, $v_g(\Delta_{\rm A}=0)=2\Omega_c^2d/\ga$ is not affected by the specific value of $k_{\rm p}d$, as it is calculated from the trace of the Green's function matrix (i.e., it only depends on the local Green's function).

We have analyzed the problem of EIT as a possible example where important properties of the system, such as the band structure, can easily be obtained by applying the product expression for the normalized transmission. In principle, any other process that respects the mathematical structure of Eq.~(\ref{transmission}) could be subject to such an analysis.

\section{Experimental perspectives}
\begin{figure*}
\centerline{\includegraphics[width=\textwidth]{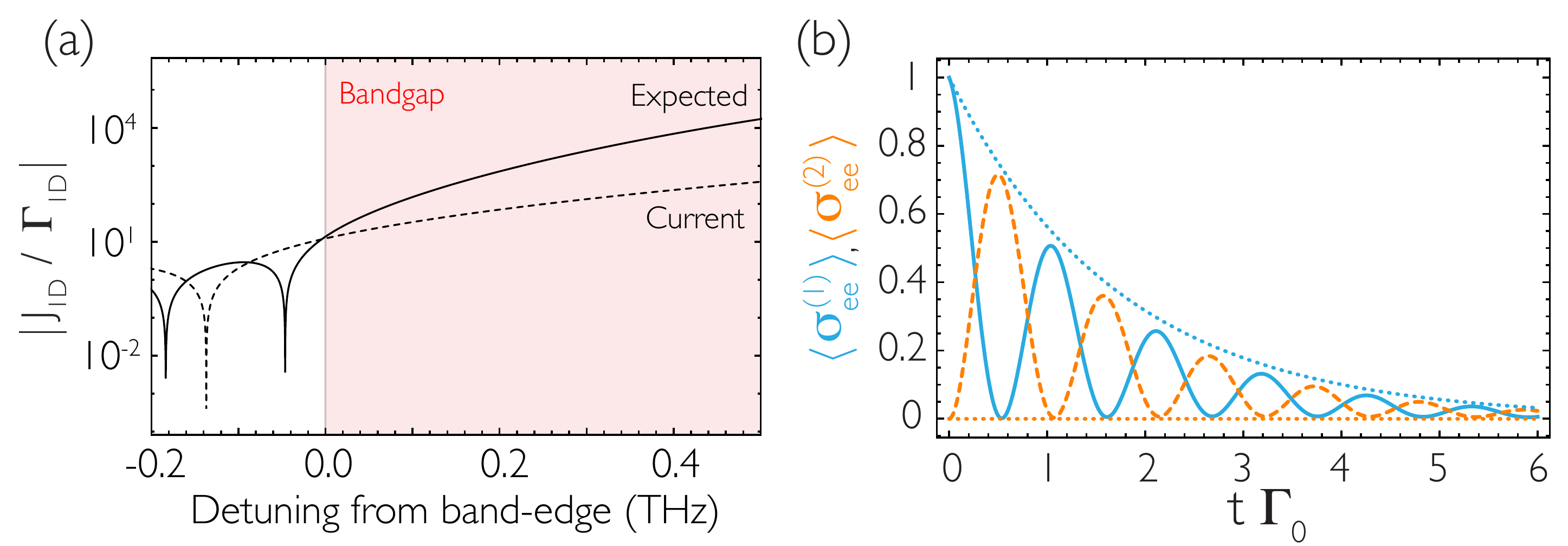}}
\caption{(a) Magnitude of the ratio between the coherent and dissipative couplings through the guided mode of an alligator PCW \cite{HGA16}, and a slot PCW, as discussed in text. The dashed line shows the ratio as given in Fig.4 of Ref.~\cite{HGA16}, and the continuous curve represents the expected ratio that could be achieved within the next years (see text for more details). (b) Evolution of the excited state population of atom 1 (blue curve) and 2 (orange dashed curve) after fully inverting atom 1 at the initial time. The resonance frequency of the atoms lies in the bandgap of the photonic crystal (at a frequency $\nu_{\rm BG}=20$~GHz from the band edge), with the atoms placed at successive even antinodes [continuous curve in (a)].  The dotted lines represent the non-interacting scenario, where the off-diagonal terms of $\mathfrak{g}$ are zero. The spin exchange and decay rates are chosen to be $J_{\rm 1D}^{\rm max}(\nu_{\rm BG})=-3\Gamma_0$, $\Gamma_{\rm 1D}^{\rm max} (\nu_{\rm BG})=0.15\Gamma_0$, and $\Gamma'=0.5\Gamma_0$. The lattice constant is $a=370$~nm and the range of interaction is $\kappa_x^{-1}= 80a$.} \label{Fig4b}
\end{figure*}

In a recent experiment \cite{HGA16}, the authors have observed signatures  of collective atom-light interactions in the transmission spectra of atoms coupled to an alligator photonic crystal waveguide. They have recorded these spectra for various frequencies around the band edge of the PCW, exploring different physical regimes. Outside the bandgap, due to the finite size of the PCW, they observe the formation of a low-finesse cavity mode [as shown in Fig.~3(a) of Ref.~\cite{HGA16}, at a frequency $\nu_1$]. At resonance with this cavity mode, the dissipative single-atom coupling to the structure for an atom at the peak of the Bloch function is $\Gamma^{\rm max}_{\rm 1D} (\nu_1)\simeq 1.5\Gamma_0$, as obtained from steady-state transmission lineshape measurements. The decay rate into leaky modes is $\Gamma'/\Gamma_0\simeq 1.1$, estimated from finite-difference time-domain (FDTD) numerical calculations. 

After tuning the spectral features of the PCW so that the resonance frequency of the atoms moves into the bandgap, they observe asymmetric lineshapes, revealing significant coherent coupling. Specifically at $\nu_{\rm BG}=60$~GHz inside the bandgap, the spin exchange and decay rates are $J^{\rm max}_{\rm 1D}(\nu_{\rm BG})/\Gamma_0\simeq -0.2$ and $\Gamma^{\rm max}_{\rm 1D}(\nu_{\rm BG})/\Gamma_0\simeq 0.01$, respectively. Due to the evanescent character of the field in the bandgap, the interaction range is finite, and at $\nu_{\rm BG}$ its value is $\kappa_x^{-1}\simeq 80a$, being $a=370$~nm the lattice constant of the alligator PCW. While this experiment constitutes the first observation of more than one emitter interacting through the guided modes around the band edge of a PCW, the values of $J_{\rm 1D}$ and $\Gamma_{\rm 1D}$ are not yet good enough to observe further signatures of atom-atom interactions such as time-dependent spin exchange. Nevertheless, we expect that near-term advances of the current set up will yield dramatic improvements on these rates, opening the door to exploring exciting collective atomic phenomena.

In particular, instead of using an alligator PCW, one can employ a slot photonic crystal waveguide \cite{LMS15,PKZ15}, i.e. a quasi-1D waveguide embedded in a 2D photonic crystal. This structure would be advantageous due to several reasons. First of all, it inhibits atomic emission into non-guided modes due to the surrounding 2D photonic bandgap that reduces the modes into which the atom can radiate. Absent inhomogeneous broadening, early simulations demonstrate that it is possible to achieve a very small non-guided decay rate, i.e. $\Gamma'\simeq 0.5\Gamma_0$. Moreover, one can engineer flatter bands,  which leads to an increase of the group index of $n_{\rm g}\simeq 30$ near the band-edge (three times larger than that of the current alligator), according to FDTD simulations. Then, both $J_{\rm 1D}$ and $\Gamma_{\rm 1D}$ would experience a three-fold increase. Finally, by trapping the atoms at the center of the nanostructure, in between the two slots and not above as it is currently done, we have estimated that $J_{\rm 1D}$ and $\Gamma_{\rm 1D}$ would be five times larger. Summarizing, we project $\Gamma^{\rm max}_{\rm 1D}(\nu_1)/\Gamma'\simeq 44$ at the first cavity resonance. This yields the values of $J^{\rm max}_{\rm 1D}(\nu_{\rm BG})/\Gamma'\simeq -6$, $\Gamma^{\rm max}_{\rm 1D}(\nu_{\rm BG})/\Gamma'\simeq0.3$, and $\Gamma'/\Gamma_0\simeq0.5$ for a detuning from the band edge $\nu_{\rm BG}=20$~GHz, where the range of interaction is $\kappa_x^{-1}\simeq 80a$.

Figure~\ref{Fig4b}(a) compares the ratio $|J_{\rm 1D}/\Gamma_{\rm 1D}|$ between the coherent and dissipative guided-mode rates for the current alligator PCW (dashed line) and the described slot PCW (continuous line). The improved ratio for the later structure can already be observed at frequencies just beyond the band-edge, and becomes $|J_{\rm 1D}/\Gamma_{\rm 1D}|\simeq10^4$ at a detuning of $0.5$~THz from the band-edge. Another signature of collective behavior is represented in Fig.~\ref{Fig4b}(b), which shows the evolution of the excited state populations of two atoms placed at successive even antinodes (continuous and dashed curves), after initially inverting one of them. The atoms interact through the guided modes of the already described slot PCW, and their resonance frequency lies inside the bandgap, at the frequency for which the interaction range is $\kappa_x^{-1}\simeq 80a$. The dotted lines show the expected result for non interacting atoms, where the off-diagonal terms of $\mathfrak{g}$ are zero, a situation that occurs when the atoms are separated by a distance $d\gg\kappa_x^{-1}$.

To summarize, we believe that there is a bright future for experiments involving not only atoms, but also superconducting qubits interacting through the guided modes of a microwave photonic crystal. In a recent experiment, a ratio of $\Gamma_{\rm 1D}/\Gamma'=50$ has already been achieved for transmon qubits connected to a 1D coplanar microwave transmission line \cite{LFL13}. Combined with the exciting recent advances in microwave photonic crystal fabrication \cite{LH16}, we expect a next generation of experiments where many qubits interact with each other in a mostly coherent manner.

\section{Conclusion}
We have analyzed the optical response of a chain of atoms placed along a quasi-1D nanophotonic structure in terms of the classical electromagnetic Green's function. This formalism is valid in the presence of absorptive and dispersive media.

We find that the linear response of the atoms can be understood in terms of collective atomic eigenstates of the Green's function matrix $\mathfrak{g}(x_i,x_j)$ for all pairs of atoms. In particular, we have derived a closed expression for the transmission spectra that only depends on the cooperative frequency shifts and decay rates of these modes. We have shown that the transmission coefficient is a direct probe of the Green's function of the nanostructure, enabling us to determine whether the atom-light interactions are fundamentally dispersive or dissipative in character as well as to quantify the degree of cooperative interaction. We have gained insight into the interactions between atoms and quasi-1D cavities, waveguides, and photonic crystals, structures of relevance in recent experiments, as well as provided estimations of what can be observed in the near future.

The Green's function formalism provides a natural language that unifies nanophotonics and quantum optics, and our results apply not only to atoms \cite{HGA16}, but to many other quantum emitters, such as superconducting qubits \cite{LH16}, NV centers \cite{AGP11}, rare earth ions \cite{ZKM15} or quantum dots \cite{LMS15}, interacting with any kind of quasi-1D photonic structures or circuits.

\textbf{Acknowledgments --} We thank S.-P. Yu for sharing his insight on slot photonic waveguides. We are also grateful to O. J. Painter, A. Keller, M. Fang, and P. Dieterle for stimulating discussions about superconducting qubits. Funding is provided by the AFOSR QuMPASS MURI, NSF Grant PHY-1205729, the Office of Naval Research (ONR) Award N00014-16-1-2399; the ONR QOMAND MURI;  the DOD NSSEFF program, and the IQIM, an NSF Physics Frontiers Center with support of the Moore Foundation. A. A.- G. was supported by the IQIM Postdoctoral Fellowship and the Global Marie Curie Fellowship LANTERN (655701). DEC acknowledges support from Fundacio Privada Cellex Barcelona, Marie Curie CIG ATOMNANO, MINECO Severo Ochoa Grant SEV-2015-0522, ERC Starting Grant FoQAL, and CERCA Programme / Generalitat de Catalunya.

\appendix
\section{Transmission and reflection coefficients in terms of Green's functions}\label{ApA}

We begin by recalling Eq.~(\ref{fieldeig}),
\begin{align}
E^+_\beta(\rb)&=E^{+}_{\rm p,\beta}(\rb)-\sum_{\xi=1}^N \frac{\left(\ggm^T_{\alpha\beta}(\rb)\cdot\textbf{v}_{\xi}\right) \left(\textbf{v}_{\xi}^T \cdot\Eb_{\rm p,\alpha}^+\right)}{(\Delta_{\rm A}+J_{\xi,\rm 1D})+\ii(\Gamma'+\Gamma_{\xi,\rm 1D})/2},
\end{align}
which relates the $\beta$-polarization component of the field along any point of the structure with the collective atomic modes. In order to calculate the transmission spectra, we need an expression that connects the output and the input fields. To do so, let's consider that we have a dipole $p_{\rm left}$ placed to the left of the first atom of the chain, at position $\rb_{\rm left}$, which is the source of the probe field $E_{\rm p,\beta}^+$. The dipole $p_{\rm left}$ is polarized along $\hat{\beta}$, the same polarization of the guided mode field. To obtain the transmission coefficient, we evaluate the field at position $\rb_{\rm right}$, immediately to the right of the last atom of the chain. When the atoms are not present, the probe field at the left and right positions of the quasi-1D nanostructure are
\begin{subequations}
\begin{equation}
E_{\rm p,\beta}^+(\rb_{\rm left})=\mu_0\omega^2_{\rm p}\, G_{\rm 1D,\beta\beta}(\rb_{\rm left},\rb_{\rm left})\, p_{\rm left},
\end{equation}
\begin{equation}
E_{\rm p,\beta}^+(\rb_{\rm right})=\mu_0\omega^2_{\rm p}\, G_{\rm 1D,\beta\beta}(\rb_{\rm right},\rb_{\rm left}) \,p_{\rm left}.
\end{equation}
\end{subequations}\\
Then, the transmission for the system without the atoms is simply
\begin{align}
t_0(\Delta_{\rm A})=\frac{E_{\rm p,\beta}^+(\rb_{\rm right})}{E_{\rm p,\beta}^+(\rb_{\rm left})}=\frac{G_{\rm 1D,\beta\beta}(\rb_{\rm right},\rb_{\rm left})}{G_{\rm 1D,\beta\beta}(\rb_{\rm left},\rb_{\rm left})}.
\end{align}

When $N$ atoms are placed in the vicinity of the nanostructure, the field at position $\rb_{\rm right}$ is
\begin{widetext}
\begin{align}\nonumber
E^+_\beta(\rb_{\rm right})&=E_{\rm p,\beta}^+(\rb_{\rm right})-\frac{1}{g_{\beta\beta}(\rb_{\rm left},\rb_{\rm left})}\sum_{\xi=1}^N \frac{\left(\ggm^T_{\alpha\beta}(\rb_{\rm right})\cdot\textbf{v}_{\xi}\right) \left(\textbf{v}_{\xi}^T \cdot\ggm_{\alpha\beta}(\rb_{\rm left})\right)}{(\Delta_{\rm A}+J_{\xi,\rm 1D})+\ii(\Gamma'+\Gamma_{\xi,\rm 1D})/2}E_{\rm p,\beta}^+(\rb_{\rm left})\\
&=\left(t_0(\Delta_{\rm A})-\frac{1}{g_{\beta\beta}(\rb_{\rm left},\rb_{\rm left})}\sum_{\xi=1}^N \frac{\left(\ggm_{\alpha\beta}^T(\rb_{\rm right})\cdot\textbf{v}_{\xi}\right) \left(\textbf{v}_{\xi}^T \cdot\ggm_{\alpha\beta}(\rb_{\rm left})\right)}{(\Delta_{\rm A}+J_{\xi,\rm 1D})+\ii(\Gamma'+\Gamma_{\xi,\rm 1D})/2}\right)E_{\rm p,\beta}^+(\rb_{\rm left}),
\end{align}
\end{widetext}
where we have employed that the probe field at atom $\rb_j$ can be related to $E_{\rm p,\beta}^+(\rb_{\rm left})$ as 
\begin{align}\nonumber
E_{\rm p,\beta}^+(\rb_j)&=\mu_0\omega^2_{\rm p} \,G_{\rm 1D,\beta\beta}(\rb_j,\rb_{\rm left})\, p_{\rm left}\\
&=\frac{G_{\rm 1D,\beta\beta}(\rb_j,\rb_{\rm left})}{G_{\rm 1D,\beta\beta}(\rb_{\rm left},\rb_{\rm left})}E_{\rm p,\beta}^+(\rb_{\rm left}).
\end{align}
Then, the normalized transmission acquires the form given by Eq.~(\ref{tandr}a) in the main text. It is important to notice that if there is only one guided mode (or if the coupling of the atoms to the other ones can be neglected), the transmission coefficient is an scalar, i.e., by calculating the coefficient for a component of the field, we calculate it for the full guided mode. 

Let's now calculate the reflection coefficient. Without the atoms, the field at $\rb_{\rm left}$ is $E^+_\beta(\rb_{\rm left})=[1+r_0(\Delta_{\rm A})]E_{\rm p,\beta}^+(\rb_{\rm left})$. When the atoms are present, the field reads 
\begin{align}
&E_\beta^+(\rb_{\rm left})=[1+r_0(\Delta_{\rm A})]E_{\rm p,\beta}^+(\rb_{\rm left})\\\nonumber
&-\frac{E_{\rm p,\beta}^+(\rb_{\rm left})}{g_{\beta\beta}(\rb_{\rm left},\rb_{\rm left})}\sum_{\xi=1}^N \frac{\left(\ggm_{\alpha\beta}^T(\rb_{\rm left})\cdot\textbf{v}_{\xi}\right) \left(\textbf{v}_{\xi}^T \cdot\ggm_{\alpha\beta}(\rb_{\rm left})\right)}{(\Delta_{\rm A}+J_{\xi,\rm 1D})+\ii(\Gamma'+\Gamma_{\xi,\rm 1D})/2}.
\end{align}
Following similar steps as those above, we find Eq.~(\ref{tandr}b) of the main text.

\section{Derivation of Equation~(\ref{transmission}) for the transmission}\label{ApB}
We can exploit some properties of 1D systems to arrive to the closed expression for the transmission shown in Eq.~(\ref{transmission}), which only depends on the decay rates and frequency shifts of the modes, not on their spatial structure (i.e. the eigenfunctions). We first show how to derive the 1D Green's function wave equation, and how the solution is related to the full quasi-1D solution.  We start with the 3D Green's function $\GG_{\rm 1D}$ for the guided mode, which follows Eq.~(\ref{gf}). We assume that the guided modes are transverse waves that travel in the $\pm \hat{x}$ direction and are polarized along $\hat{y}$, and that the field is approximately uniform in the transverse directions. From 3D, one can in principle construct the guided modes and their dispersion relations $\omega(k)$, from which one can identify an effective dielectric constant $\epsilon_{\rm eff}(x,\omega)$ which produces the same behavior (at least within some bandwidth). The final answer that we are trying to achieve does not depend on explicit construction of $\epsilon_{\rm eff} (x,\omega)$. The result is a Helmoltz equation for the Green's function that reads
\begin{equation}
\left[  \frac{d^2}{d x^2} + \frac{\omega^2}{c^2} \epsilon_{\rm eff}(x,\omega)   \right] \tilde{G}_\text{1D}(x,x',\omega) = - \delta(x-x'),
\label{eq:1DGF}
\end{equation}
where $\tilde{G}_{\rm 1D}=AG_{\rm 1D}$, being $A$ the effective mode area. The solution for this second order linear ordinary differential equation can be expressed as the sum of the two homogeneous solutions. The Green's function can then be written in terms of the auxiliary fields $\tilde{\phi}_{L,R}(x)$, which are solutions of the homogeneous equation, as
 \begin{align}\nonumber
\tilde{G}_\text{1D}(x,x') =\frac{1}{W}&\left[\Theta(x'-x) \tilde{\phi}_L(x') \tilde{\phi}_R(x)\right.\\
+&\left.\Theta(x-x') \tilde{\phi}_R(x') \tilde{\phi}_L(x)\right],
\label{eq:GWronskian}
\end{align}
where $W$ is the Wronskian, which does not depend on the position, and is given by 
\begin{equation}
W = \tilde{\phi}_R(x')  \frac{\textrm{d} \, \tilde{\phi}_L(x') }{\textrm{d}x'}   -  \frac{\textrm{d} \,\tilde{\phi}_R(x')}{\textrm{d}x'}\tilde{\phi}_L(x').
\label{eq:}
\end{equation}
We can then recover the full Green's function between atom $i$ and atom $j$ as
\begin{align}\label{wrons}
&G_{\rm 1D} (x_i,x_j,\omega)=\frac{1}{A} \tilde{G}_{\rm 1D} (x_i,x_j,\omega)\\\nonumber
&= \left[\Theta(x_j-x_i)\phi_L(x_j) \phi_R(x_i) +\Theta(x_i-x_j) \phi_L(x_i) \phi_R(x_j)\right],
\end{align}
where $\phi_{R,L}\equiv\tilde{\phi}_{R,L}/\sqrt{AW}$. Then, the dipole-projected Green's function is
 \begin{equation}
g_{ij}=\Theta(x_j-x_i)s_{ji}+\Theta(x_i-x_j)s_{ij},
\end{equation}
where $s_{ij}=\varphi_{L}(x_i) \varphi_{R}(x_j)$, with $\varphi_{L,i}=\sqrt{\mu_0\omega_{\rm p}^2d^2/\hbar}\,\,\phi_L(x_i)$ and $\varphi_{R,j}=\sqrt{\mu_0\omega_{\rm p}^2d^2/\hbar}\,\,\phi_R(x_j)$. It is convenient to define the rank-one matrix $\mathfrak{s}={\bm \varphi}_L\otimes{\bm \varphi}_R^T$, where ${\bm \varphi}_{\{R,L\}}=(\varphi_{\{R,L\}}(x_1), ..., \varphi_{\{R,L\}}(x_N))$ is a vector of $N$ components. Let's now proceed to demonstrate Eq.~(\ref{transmission}). In terms of the eigenfunctions of $\mathfrak{g}$, the transmission is
\begin{align}\nonumber
&t(\Delta_{\rm A})/t_0(\Delta_{\rm A})=\\\nonumber
&1-\frac{1}{g(x_{\rm right},x_{\rm left})}\sum_{\xi=1}^N \frac{\left(\ggm^T(x_{\rm right})\cdot\textbf{v}_{\xi}\right) \left(\textbf{v}_{\xi}^T \cdot\ggm(x_{\rm left})\right)}{(\Delta_{\rm A}+J_{\xi,\rm 1D})+\ii(\Gamma'+\Gamma_{\xi,\rm 1D})/2}\\
&=1-\frac{1}{g(x_{\rm right},x_{\rm left})}\left(\ggm^T(x_{\rm right})\cdot\mathcal{M}^{-1}\cdot\ggm(x_{\rm left})\right),
\end{align}
where $\mathcal{M}$ is given in Eq.~(\ref{sigmam}). Since $\ggm\propto G_{\rm 1D}$, and using the expression for the Green's function in terms of the right-going and left-going field solutions [Eq.~(\ref{wrons})], we find
\begin{align}\nonumber
t(\Delta_{\rm A})/t_0(\Delta_{\rm A})&=1-{\bm \varphi}^T_R\cdot\frac{1}{\Delta_{\rm A}+\ii\Gamma'/2+\mathfrak{g}}\cdot{\bm \varphi}_L\\
&= 1-\textbf{w}^T\cdot\frac{1}{\mathbb{1}+\tilde{\mathfrak{g}}}\cdot\textbf{u},
\end{align}
where we have defined $\textbf{w}\equiv{\bm \varphi}_R/\sqrt{\Delta_{\rm A}+\ii\Gamma'/2}$, $\textbf{u}\equiv{\bm \varphi}_L/\sqrt{\Delta_{\rm A}+\ii\Gamma'/2}$, and $\tilde{\mathfrak{g}}\equiv\mathfrak{g}/(\Delta_{\rm A}+\ii\Gamma'/2)$. By the matrix determinant lemma \cite{HJ13}, we know that for a invertible matrix $\textbf{A}$ and a pair of vectors $\textbf{u},\textbf{w}$, we can write $\text{det}(\textbf{A}+\textbf{u}\otimes\textbf{w}^T)=\text{det}(\textbf{A})\left(1+\textbf{w}^T\cdot\textbf{A}^{-1}\cdot\textbf{u}\right)$. Choosing $\textbf{A}=-(\mathbb{1}+\tilde{\mathfrak{g}})$, we find
\begin{align}\nonumber
t(\Delta_{\rm A})/t_0(\Delta_{\rm A})&=\frac{\text{det}(\mathbb{1}+\tilde{\mathfrak{g}}-\textbf{u}\otimes\textbf{w}^T)}{\text{det}(\mathbb{1}+\tilde{\mathfrak{g}})}\\
&=\frac{\text{det}((\Delta_{\rm A}+\ii\Gamma'/2)\mathbb{1}+\mathfrak{g}-\mathfrak{s})}{\text{det}((\Delta_{\rm A}+\ii\Gamma'/2)\mathbb{1}+\mathfrak{g})}.
\end{align}
Since $(\Delta_{\rm A}+\ii\Gamma'/2)\mathbb{1}+\mathfrak{g}-\mathfrak{s}$ is a triangular matrix with $(\Delta_{\rm A}+\ii\Gamma'/2)$ in the diagonal entries, and the determinant of a triangular matrix is the product of the diagonal entries, we find $\text{det}((\Delta_{\rm A}+\ii\Gamma'/2)\mathbb{1}+\mathfrak{g}-\mathfrak{s})=(\Delta_{\rm A}+\ii\Gamma'/2)^N$, which yields
\begin{align}
t(\Delta_{\rm A})/t_0(\Delta_{\rm A})&=\frac{(\Delta_{\rm A}+\ii\Gamma'/2)^N}{\text{det}((\Delta_{\rm A}+\ii\Gamma'/2)\mathbb{1}+\mathfrak{g})}\\\nonumber
&=\prod_{\xi=1}^N\,\frac{\Delta_{\rm A}+\ii\Gamma'/2}{(\Delta_{\rm A}+J_{\xi,\rm 1D})+\ii(\Gamma'+\Gamma_{\xi,\rm 1D})/2},
\end{align}
as the determinant of a matrix is the product of its eigenvalues. The above expression is precisely Eq.~(\ref{transmission}). To the best of our knowledge, it is not possible to obtain a simplified expression for the reflection coefficient.

\section{Non-Markovian effects: colored reservoirs}\label{ApC}
 \begin{figure}
\centerline{\includegraphics[width=\linewidth]{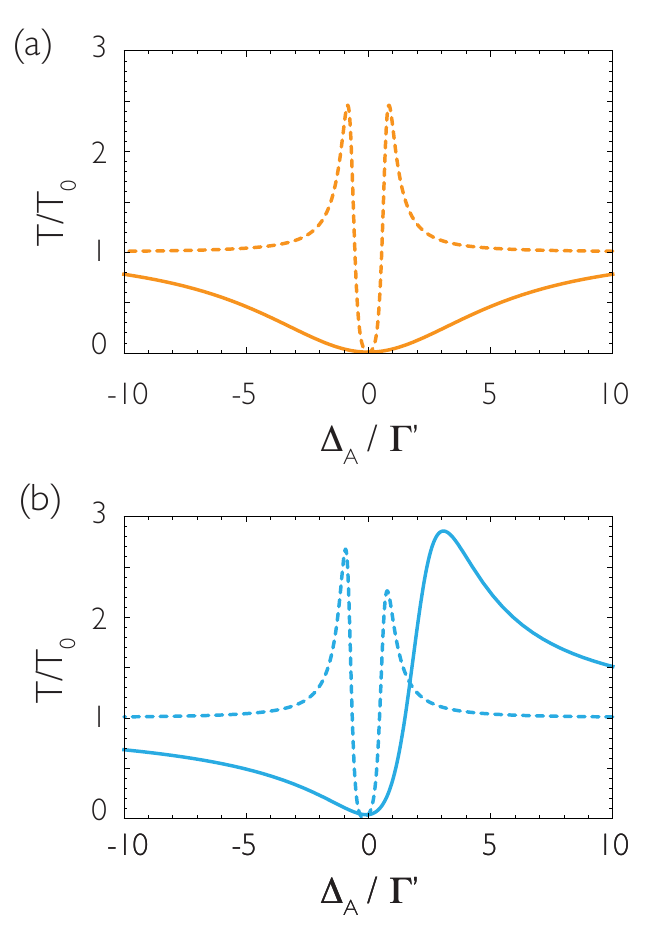}}
\caption{Markovian (continuous curves) and non-Markovian (dashed lines) normalized transmission spectra in the dissipative (a) and dispersive (b) regimes of a cavity containing 10 atoms located at even numbered antinodes of the cavity field. For the dashed lines, the cavity linewidth is $\kappa_{\rm c}=0.2\Gamma'$, whereas for the continuous curves, $\kappa_{\rm c}=1000\Gamma'$. The upper plot is evaluated at the cavity resonance, whereas the lower plot is evaluated at a detuning from the cavity resonance $\Delta_{\rm c}=\kappa_c$. For all figures, $\Gamma_{\rm 1D}=\Gamma'$.} \label{Fig5}
\end{figure}
The Markov approximation has been employed in the analysis carried out in the main text, as the Green's function of the nanostructures are considered to be constant over frequency ranges larger than the linewidth of the atoms.

If that is not the case, it is not possible to find simple expressions for the Hamiltonian and Lindblad terms for the atomic density matrix. However, the expressions for the transmission and reflection coefficients are valid even when the spectral variation of the Green's function occurs within frequency intervals comparable to and smaller than the atomic linewidth. This fact might not be surprising, as in the low saturation limit, atoms behave as classical dipoles, and an equation for the transmission and reflection coefficients identical to those in Eqs.~(\ref{tandr}a-b) can be found for classical emitters, without resorting to Markov's approximation.

Following Refs.~\cite{WSL04,YVR09}, one finds the following equations for the field and atomic coherence operators in the frequency domain
\begin{align}
&\hat{\Eb}^+(\rb,\omega)=\hat{\Eb}_{\rm p}^+(\rb,\omega)+\mu_0 \omega^2 \sum_{j=1}^N \GG(\rb,\rb_j,\omega)\cdot\db\,\hge^j(\omega),\\
&\hge^j(\omega)=-\frac{1}{\hbar\Delta_{\rm A}}\db^*\cdot\hat{\Eb}^+(\rb,\omega).
\end{align}
By plugging the expression for the field into the equation for the atomic coherence operators, one obtains Eq.~(\ref{sigmam}), but for the operators in the frequency domain and without any Markovian impositions on the Green's function. Then, one finds the expressions for the transmission and reflection coefficients of Eqs.~(\ref{tandr}a-b), but where now the eigenvalues are frequency-dependent and not constant quantities. 

Therefore, these expressions could be employed to understand the spectrum of a very recent experiment of a superconducting qubit in the bandgap of a photonic crystal waveguide, where the Green's function of the structure varies significantly within the linewidth of the qubit \cite{LH16}. 
 Figure~\ref{Fig5} shows Markovian (continuous lines) and non-Markovian transmission spectra (dashed lines) in the dissipative and dispersive regimes of a cavity in the strong coupling limit. The non-Markovian curves show the typically encountered Rabi splitting, which results in fractional decay of the population and oscillations in time domain, in contrast with the exponential decay of the Markovian case.

\bibliographystyle{unsrt}
\bibliography{refs}
\end{document}